\documentclass[11pt]{article}

\usepackage{amssymb,latexsym, amsmath}
\usepackage{graphicx}

\makeatletter \@addtoreset{figure}{section}
\def\thefigure{\thesection.\@arabic\c@figure} \def\fps@figure{h, t}
\@addtoreset{table}{bsection}
\def\thetable{\thesection.\@arabic\c@table} \def\fps@table{h, t}
\@addtoreset{equation}{section}

\makeatother

\topmargin by -\headsep \textheight 8.9in \oddsidemargin 0pt
\evensidemargin \oddsidemargin \marginparwidth 0.5in \textwidth
6.5 truein

\newtheorem{thm}{Theorem}[section]

\begin{document}

\title{Wave Solutions of Evolution
Equations and Hamiltonian Flows on Nonlinear Subvarieties of Generalized
Jacobians
\footnote{PACS numbers 03.40.Gc, 11.10.Ef, 68.10.-m, AMS
Subject Classification 58F07, 70H99, 76B15}}
\author {Mark S. Alber
\thanks{Research partially supported by NSF grant DMS 9626672.} \\
Department of Mathematics\\ University of Notre Dame\\ Notre Dame,
IN 46556 \\ {\footnotesize Mark S. Alber.1@nd.edu} \and Yuri N.
Fedorov\\ Department of Mathematics and Mechanics\\ Moscow
Lomonosov University\\ {\footnotesize fedorov@mech.math.msu.su }
 \\}
\date{December 14, 1999}

\maketitle

\newpage

\begin{abstract}
The algebraic-geometric approach is extended to study solutions of
$N$-component systems associated with the energy dependent
Schr\H{o}dinger operators having potentials with {\it poles} in
the spectral parameter, in connection with Hamiltonian flows on
nonlinear subvariaties of Jacobi varieties. The systems under
study include the shallow water  equation and Dym type equation.
The classes of solutions are described in terms of theta-functions
and their singular limits by using new parameterizations. A
qualitative description of real valued solutions is provided.
\end{abstract}

\tableofcontents

\section{Introduction.}
\setcounter{equation}{0} The quasi-periodic solutions of most
classical integrable PDEs can be obtained using the inverse
scattering transform (IST)
 (see, for example, Ablowitz and Segur [1981], Newell [1985] and
Ablowitz and Clarkson  [1991]). This is done by establishing a
connection with an isospectral eigenvalue problem for an associated
Schr\"{o}dinger operator.

The solution of nonlinear evolution equations using
algebraic-geometric techniques was initially developed to handle
$N$-phase wave trains. This approach can be summarized as follows.
By using the trace formula, families of quasi-periodic and soliton
solutions are associated with Hamiltonian flows on finite
dimensional phase spaces. These flows are described by using so
called $\mu$-variable representations leading to an Abel--Jacobi
mapping which include holomorphic and, in some cases, meromorphic
differentials (see amongst others, Ablowitz and Ma [1981],
Dubrovin [1981], Ercolani [1989] and Alber and Alber [1985]). Then
the mapping is inverted in terms of Riemann theta-functions and
their singular limits. Many well-known nonlinear equations such as
KdV, sine-Gordon, focusing and defocusing nonlinear
Schr\H{o}dinger equations, which describe a wide variety of
important phenomena in physics, optics, biology and engineering,
were studied by using this approach.

Recently special attention was given to the
shallow water (SW) equation derived in Cammasa and Holm [1993]
in the context of the Hamiltonian structure,
\begin{equation}
\label{CH-sw-eqn}
U_t+3UU_x = U_{xxt}+2U_x U_{xx}+U U_{xxx}-2\kappa U_x\,,
\end{equation}
and the Dym type equation (see  Cewen [1990], Hunter and Zheng
[1994] and Alber {\it et al.} [1994, 1995])
\begin{equation}
\label{dym-sw-eqn}
U_{xxt}+2U_x U_{xx}+U U_{xxx}-2\kappa U_x=0, \qquad \kappa={\rm const}.
\end{equation}
Camassa and Holm [1993] described classes of $n$-peakon
soliton-type solutions for an integrable (SW) equation
(\ref{CH-sw-eqn}). In particular, they obtained a system of
completely integrable Hamiltonian equations for the locations of
the ``peaks'' of the solution, the points at which its spatial
derivative changes sign. In other words, each peakon solution can
be associated with a mechanical system of moving particles.
Calogero [1995] and Calogero and Francoise [1996] further extended
the class of mechanical systems of this type.

The problem of describing complex traveling wave and
quasi-periodic solutions of the equations (\ref{CH-sw-eqn}) and
(\ref{dym-sw-eqn}) can be reduced to solving finite-dimensional
Hamiltonian systems on symmetric products of hyperelliptic curves.
Namely, according to Alber {\it et al} [1994,1995,1999], such
solutions can be represented in the following form $$
U(x,t)=\mu_1+\cdots+\mu_g -M, $$ where $g$ is a positive integer,
$M$ is a constant and the evolution of the variables $\mu$ is
given by the equations
\begin{equation}
\label{A-J-g}
\sum_{i=1}^g \frac{\mu_i^{k}\,{\rm d}\mu_i}{2\sqrt{R(\mu_i)}}
= \left\{ \begin{array}{lll}   0 & \mbox{$k=1,\dots,g-2,$} \\
                            {\rm d} t & \mbox{$k=g-1,$} \\
                            {\rm d} x & \mbox{$k=g.$} \end{array} \right.
\end{equation}
Here $R(\mu)$ is a polynomial of degree $2g+2$ (for shallow water equation
(\ref{CH-sw-eqn}))
or $2g+1$ (for the Dym type equation (\ref{dym-sw-eqn})). Also $M=0$ for
the Dym type
equation.

In contrast to the finite-dimensional reductions of such equations as KdV
and sine-Gordon
equations,
system (\ref{A-J-g}) contains a meromorphic differential, also the
number of holomorphic differentials
is less than the genus $g$ of the corresponding hyperelliptic curve: $W^2 =
R(\mu)$.
This implies that the problem of inversion (\ref{A-J-g}) can not be solved
in terms of meromorphic functions of $x$ and $t$.
Examples of such equations  arise in
several problems of mechanics. These were considered in Vanhaecke [1995] and
Abenda and Fedorov [1999],
where a connection was established with the flows on nonlinear
subvarieties of hyperelliptic Jacobian varieties, so-called ${\it strata}$.
In Alber {\it et al.} [1997] a whole class of
$N$-component systems with
poles was shown to be integrable by reducing them to similar nonstandard
inversion problems which contained meromorphic differentials. Therefore
$N$-component systems can be overdetermined, implying that
the genus of the spectral curve can be higher then the number
of $\mu$-variables.

$N$-component systems can be briefly described as follows.
For the KdV equation,
the spectral parameter appears linearly in the potential of the
corresponding Schr\"{o}dinger equation: $V = u - \lambda$, in the context
of the IST method. In
contrast, Antonowicz and Fordy [1987a,b, 1988, 1989] and Antonowicz, Fordy
and Liu
[1991]
investigated potentials with poles in the spectral parameter for what they
refer to as {\it energy dependent} Schr\"{o}dinger operators connected to
certain systems of evolution equations.
Specifically, they obtained multi-Hamiltonian structures for
$N$-component integrable systems of equations related to the following
isospectral eigenvalue problem:
\begin{equation} \label{iegenvalue} {\displaystyle L \psi = \left (
\frac{\partial^2}{\partial x^2} + \frac{V}{K} \right ) } \psi =  0\ ,
\end{equation}
\begin{equation} \label{kv} K = \sum_{j = 0}^{M} k_j \lambda^{j}; \;\;\; V =
\sum_{j = 0}^{N} v_j (x,t) \lambda^{j}\ ,
\end{equation} where the $k_j$ are constants and the $v_j (x,t)$ are functions
of the variable $x$, the parameter $t$ and the spectral parameter
$\lambda$ is complex. This includes the coupled KdV and Dym systems.

In Alber {\it et al} [1994, 1995, 2000a], the presence of a pole
in the potential was shown to be necessary for the existence of
weak billiard solutions of nonlinear equations. Billiard solutions
of nonlinear PDE's have been related to finite-dimensional
integrable dynamical systems with reflections including
ellipsoidal Birkhoff billiards. It turned out that the existence
of billiard solutions and the presence of monodromy effects is a
specific feature of the whole class of $N$-component systems with
poles (see Alber {\it et al.} [1997]).

Quasi-periodic solutions of the Dym equation were studied in
Dmitrieva [1993a] and Novikov [1999] by using a connection with
KdV equation and introducing additional phase functions. Soliton
solutions of the Dym type equation were studied in Dmitrieva
[1993b]. Periodic solutions of the shallow water equation were
discussed in McKean and Constantin [1999]. Beals {\it et al.}
[1999, 2000] used Stieltjes theorem on continued fractions and the
classical momemt problem for studing multi-peakon solutions of the
(SW) equation.

The main goal of this paper is to describe explicit formulae in
terms of theta-functions and their singular limits for the
solutions to the shallow water equation (\ref{CH-sw-eqn}). We also
explain the role of the mysterious phase functions used by
Dmitrieva [1993a] when studying SW equation and equation of the
Dym type. This phase is present for the whole class of WKI
hierarchy. (see Wadati {\it et al.} [1979]).

In the present paper the traveling wave, soliton, peakon, cuspon
and quasi-periodic
 solutions are  considered. Usually in the
case of integrable evolution equations quasi-periodic flows are
liberalized on the Jacobi varieties. In this paper we show that in
the case of N-component systems with poles the $x$- and $t$-flows
take place on {\it nonlinear}  subvarieties (strata) of
generalized (noncompact) Jacobians. For this reason, the term
``liberalization" is no longer applicable here. This makes the
above nonlinear equations quite different from such well known
equations as KdV, sine-Gordon and nonlinear Schr\H{o}dinger
equations. For the sake of clarity, in this paper we start with
solutions related to  (hyper)elliptic curves of at most genus 2.
The case of arbitrary genus is only notationally more complicated
and we provide complete formulae. In addition, we give a complete
classification of real bounded solutions $U(x,t)$ in the above
cases and provide corresponding plots.

Notice that the complex geometry of the traveling wave solutions,
cusp and peakon solutions was previously studied in Alber {\it et
al} [1994, 1995, 2000] in connection with geodesic flows with
reflections on Riemannian manifolds  and in Li and Olver [1998]
from the point of view of singularity analysis.

\medskip

\paragraph{The Contents of the Paper.}
In Section 2 we demonstrate the main difference between  the
nonlinear SW equation and the KdV equation from the point of view
of the algebraic-geometric approach by obtaining traveling wave
solutions as a result of inverting elliptic integrals of the
second and third kind. Here we express the amplitude $U$ and the
phase $X$ as meromorphic functions of an auxiliary variable
parameterizing the elliptic curve.

In Section 3 we apply different singular limits to the problem of
inversion resulting in formulae for different periodic and
solitary solutions. In particular, peakon solutions are obtained
as limits of the traveling wave solutions and are related to
various singularizations of the elliptic curves.

Section 4 provides explicit expressions in terms of
theta-functions for the {\it stationary} quasi-periodic solutions.
This is done by using new complex parametrizations on the  related
associated hyperelliptic curve of genus 2.

In Section 5 we find
time-dependent solutions by integrating and inverting equations (\ref{A-J-g})
in the genus 2 case. We show that these equations can be extended to a
standard
Abel--Jacobi mapping
of a symmetric product of the hyperelliptic curve to its {\it generalized}
Jacobian. The original system (\ref{A-J-g}) then defines a mapping onto a
2-dimensional nonlinear stratum of the Jacobian, a generalized theta-divisor,
where the dynamics actually takes place. By fixing $t$ in the expression
for the
solution in terms of theta-functions, we then recover the stationary
quasi-periodic
solutions obtained in Section 4.

Section 6 contains qualitative analysis of real bounded solutions for the
case when the Weierstrass points of the spectral curve are real.

In a forthcoming paper  we will consider different singular limits
of the quasi-periodic solutions when the spectral curve becomes
singular and its arithmetic genus drops to zero. The solutions are
then expressed in terms of purely exponential tau-functions and,
in the real bounded case, they describe, in particular, a
quasi-periodic train of peakons tending to a periodic one at
infinity.

\section{Traveling Wave Solutions.}
\setcounter{equation}{0}

The main difference between  the nonlinear SW or Dym
equations and the KdV equation from the point of view of the
algebraic-geometric approach
can be demonstrated already on the level of traveling wave solutions.

The
traveling wave solutions of the KdV equation are obtained
by inversion of an elliptic integral of the
{\it first} kind, i.e., the integral of
a holomorphic differential, which results in a meromorphic function.

In contrast to this, after substituting  $U(x,t)=\lambda (x-ct)$ into the Dym
type equation (\ref{dym-sw-eqn}) and using a simple transformation (see Alber
{\it et al} [1995]), we arrive at the problem of inversion of the following
Abelian integral of the {\it second} kind
\begin{equation}
\label{abel}
\int_{U_0}^U \sqrt{\frac{\lambda-c}{\kappa(\lambda-a_1)(\lambda-a_2)}}d\lambda
=\int_{U_0}^U
\frac{\lambda-c}{\sqrt{\kappa}\sqrt{R_3(\lambda)}}\,d\lambda=x-ct=X,
\end{equation}
defined on the elliptic curve ${\cal E}=\{ W^2=R_3(\lambda)\}$, where
$$
R_3(\lambda)=(\lambda-a_1)(\lambda-a_2)(\lambda-c)
$$
and $U_0$ being a constant.
Here we suppose that the roots of $R_3(\lambda)$ are distinct.
The differential in (\ref{abel}) has a double pole at infinity
$\lambda=\infty$
and a double zero at the Weierstrass point $\lambda=c$ on ${\cal E}$.
It follows that the complex inverse function $U(X)$ must have two independent
periods on $\Bbb C$, the periods of the differential along two nontrivial
homology cycles on ${\cal E}$. On the other hand, $U(X)$ blows up only at
$X=\infty$.
There are no meromorphic functions with such properties (see e.g.
Markushevich [1977]). Moreover, because of the double zero of the
differential,
the solution $U(X)$ has moving branch points of the form
\begin{equation}
\label{branching}
U-c=O((X-X_0)^{2/3}).
\end{equation}
One can show that $U(X)$ has infinitely many such branch points.

To deal with this we describe the complex function $U(X)$ in a new
parametric form by
introducing
 a new complex variable $u$ which gives a parameterization of the
curve ${\cal E}$:
\begin{equation}
\label{vr}
\int_{\lambda}^{\infty} \frac{d\lambda}{2\sqrt{R_3(\lambda)}}=u.
\end{equation}
Then, according to the theory of elliptic functions,
\begin{equation}
\lambda(u)=\wp (u)+\Delta=-\frac{d^2}{du^2}\log\theta_{11}(z)+{\rm const},
\quad
z=2\pi i\, u/\omega_3
\end{equation}
where $\wp(u)$ is the elliptic Weierstrass
function with periods $2\omega_1$ and $2\omega_3$ depending on the
coefficients
of $R_3(\lambda)$ and $\Delta=(a_1+a_2+c)/3$.
The $\theta_{11}(z)$ is the quasi-periodic Riemann theta-function
which vanishes at the points of the period lattice \\
$\Lambda=\{z=2\pi i\Bbb Z+{\bf B}\Bbb Z\}$,  ${\bf B}=\pi i\omega_1/\omega_3$
(see, e.g. Dubrovin [1981]):
$$
\theta_{11}(z)=\sum_{M\in {\Bbb Z}}\exp
\left(\frac12 {\bf B}(M+1/2)^2+(M+1/2)(z+\pi i)\right).
$$
Here we use two variables $u,z$ because of different normalizations of
(quasi)-periods of $\wp$ and $\theta$.
The curve ${\cal E}$ can be identified with the factor $\Bbb C/\Lambda$.
Now the integral (\ref{abel}) can be transformed as follows
\begin{equation}
\label{z}
\int_{u_0}^{u} (\wp (u)+\Delta-c)\,du=\sqrt{\kappa}X+{\rm const}, \quad
u_0={\rm const},
\end{equation}
which yields
\begin{equation}
\label{z1}
\sqrt{\kappa}X+{\rm const}
=\zeta(u)+(\Delta-c) u=\frac d{dz}\log\theta_{11}(z)+\Delta'z, \qquad
\Delta'={\rm const},
\end{equation}
where $\zeta(u)$ is the Riemann zeta-function with parameters
$\eta_1$ and $\eta_3$ such that for any $u\in\Bbb C$
$$
\zeta(u)=\int^{\infty}_{u}{\wp}(u)\,du, \quad
\zeta(u+2\omega_1)=\zeta(u)+2\eta_1, \quad
\zeta(u+2\omega_3)=\zeta(u)+2\eta_3.
$$
The constants $2\eta_1, 2\eta_3$ are interpreted as the
periods of the differential
$$
\lambda\,d\lambda/2\sqrt{R_3(\lambda)}=\lambda(u)\,du.
$$
Then, in view of (\ref{z1}),
$$
\sqrt{\kappa} X(u+2\omega_1)=\sqrt{\kappa} X(u)+2\eta_1+2(\Delta-c)\omega_1,
$$
$$
\sqrt{\kappa} X(u+2\omega_3)=\sqrt{\kappa} X(u)+2\eta_3+2(\Delta-c)\omega_3.
$$
\vspace{3mm}

\noindent {\it Thus we have expressed the amplitude $U=\lambda$
and the phase $X$ in terms of the auxiliary complex variable $u$}.
\vspace{3mm}

\noindent
In what follows we study  real solutions $U(X)=\lambda(u(X))$ which
correspond to the case
of all roots of the polynomial $R_3(\lambda)$ being real.
We choose the half-period $\omega_1$ to be real and $\omega_3$
to be purely imaginary.
The existence   of branch points of
$U(X)$ (see (\ref{branching})) implies
that the real solutions may have cusps.
To demonstrate this  we consider two different cases.

\paragraph{Case 1: $\kappa >0$, $a_1 < a_2 < c<\infty$.}
According to the theory of elliptic functions, parameterization
(\ref{vr}) yields that
\begin{equation}
\lambda (\omega_1)=c, \;\; \lambda (\omega_3)=a_1, \;\;
\lambda (\omega_1+\omega_3)=a_2, \;\;\lambda (0)=\infty.
\end{equation}
Along the real axis $\Re u$ and the line $\{u=\omega_3+u'|
u'\in{\Bbb R}\}$, $\lambda(u)$ is real valued. Since we are
interested in nonsingular traveling wave solutions and $\lambda(u)
$ has a pole at the origin, let us consider $\lambda
(\omega_3+u')$ as a function of $u'$. It is periodic with period
$2\omega_3$ and takes values in the interval $[a_1,a_2]$. The
differentials in (\ref{abel}) and (\ref{z}) do not have poles or
zero's on the line $\{u=\omega_3+u'\}$. Hence, $X(\omega_3+u')$
and the inverse $u'(X)$ are monotonic functions. As a result, due
to (\ref{z1}), the composition function $U(X)=\lambda
(\omega_3+u'(X))$ is a regular periodic function with period
$(2\eta_3+2(\Delta-c)\omega_3)/\sqrt{\kappa}$, and its graph is
merely a distortion of that of $\lambda (\omega_3+u')$  .

\paragraph{Case 2: $\kappa >0, a_1 < c < a_2 < \infty$.}
Here we have
\begin{equation}
\lambda (\omega_1)=a_2, \;\; \lambda (\omega_3)=a_1, \;\;
\lambda (\omega_1+\omega_3)=c, \;\;\lambda (0)=\infty.
\end{equation}
Again, $\lambda(u)$ is real valued along the real axis $\Re u$ and the line
$\{u=\omega_3+u'| u'\in{\Bbb R}\}$.
Since $\lambda (u)$ has a pole along the real axis, we again consider only
the real function $\lambda(\omega_1+u')$, $u'\in\Bbb R$ which is
$2\omega_3$-periodic and now takes values in the
interval $[c,a_2]$. In this case, the differential (\ref{z}) has
a {\it double} zero at $u=\omega_1+\omega_3$ ($u'=\omega_1$), where
the derivative $d\lambda /du$ has a simple zero. As a result, the derivative
$dU/dX=d\lambda /du\cdot du/dX$ blows up for the corresponding value of $X$,
which implies that the graph of the function $U(X)$ has cusps with periodicity
$(2\eta_3+2(\Delta-c)\omega_3)/\sqrt{\kappa}$   .
This phenomenon was first detected in Alber {\it et al} [1994,1995].

\paragraph{SW Equation.} Traveling wave solutions for the
shallow water equation have a similar description. Substituting $U(x,t)=\mu
(x-ct)$ into
(\ref{CH-sw-eqn}), we obtain an Abelian integral of the third kind
\begin{equation}
\label{abel2}
\int_{U_0}^U \sqrt{\frac{\mu-c}
{\kappa(\mu-a_1)(\mu-a_2)(\mu-a_3)}}\,d\mu
=\int_{U_0}^U \frac{\mu-c}{\sqrt{\kappa}\sqrt{R_4(\mu)}}\,d\mu =x-ct=X,
\end{equation}
defined on the even order elliptic curve $$ {\tilde {\cal
E}}=\{W^2=R_4(\mu)\}, \quad
R_4(\mu)=(\mu-a_1)(\mu-a_2)(\mu-a_3)(\mu-c), $$ where the roots of
$R_4$ are supposed to be distinct. The differential in
(\ref{abel2}) has a pair of simple poles at the infinite points
$\infty_-,\infty_+$ on ${\tilde {\cal E}}$. Then the complex
inverse function $U(X)$ must have three independent periods on
$\Bbb C$: the periods of the differential along two nontrivial
homology cycles on $\tilde{\cal E}$ and along a homology zero
cycle around one of the infinite points. A meromorphic function
with such a property does not exist.

Let $A,B$ be canonically conjugated cycles on $\tilde {\cal E}$
and $\bar\omega$ be the normalized holomorphic differential $$
\bar{\omega}=\frac{d\mu}{\Pi\sqrt{R_4(\mu)}}, $$ where the
multiplier $\Pi$ is chosen from the condition
$\oint_A\bar{\omega}=2\pi i$. Introduce a new variable $z$
parameterizing ${\tilde {\cal E}}$ as follows
\begin{equation}
\label{diff-3-kind}
\int_{\mu}^{c} \frac{d\mu}{\Pi\sqrt{R_4(\mu)}}=z,
\end{equation}
where $a_1,a_2$ and $c$ denote the corresponding Weierstrass (branch) points on
$\tilde {\cal E}$. Then we get the expression
\begin{equation}
\label{U-SW}
\mu(z)=\rho-\frac
d{dz}\log\frac{\theta[\delta](z-q/2)}{\theta[\delta](z+q/2)},
\end{equation}
$$
\rho=\oint_A\mu\, \bar{\omega}, \quad
q=2\int^{\infty_+}_{c} \frac{d\mu}{\Pi\sqrt{R_4(\mu)}}.
$$
Thus $\mu(z)$ is an elliptic function with periods $2\pi i$,
${\bf B}=\oint_B\bar{\omega}$.
The integral (\ref{abel2}) takes the form
\begin{equation}
\label{parX}
\sqrt{\kappa}X=\int_{z_0}^{z} (\mu(z)-c)\,dz=
(\rho-c)z-\log\frac{\theta_{11}(z-q/2)}{\theta_{11}(z+q/2)}+{\rm const}.
\end{equation}
As a result, we have expressed the function $U=\mu$ and its argument $X$ in
terms of
$z$.

Since the differential in (\ref{abel2}) has
a double zero at the Weierstrass point $\mu=c$, similar to the case of the Dym
equation, the inversion  of (\ref{abel2}) yields $U(X)$ with infinitely
many branch
points of the form (\ref{branching}). This  results in   real
solutions having cusps.

 Now we describe 2 different types of real traveling wave
solutions of (\ref{CH-sw-eqn}) assuming that $\kappa>0$ and all roots
of $R_4(\mu)$ are real. In this case the period ${\bf B}$ is real as well.

\paragraph{Case 1.} $a_1<a_2<a_3<c$. According to the parameterization
(\ref{diff-3-kind}), we have
$$ q\in{\Bbb R}, \quad \Pi=\int_{a_1}^{a_2} \frac{d\mu}{\sqrt{R_4(\mu)}},
\quad {\bf B}=2\int_{a_3}^{c} \frac{d\mu}{\Pi\sqrt{R_4(\mu)}},
\quad \mu(q/2)=U(-q/2)=\infty, $$
$$
\mu(0)=c,\quad  \mu({\bf B}/2)=a_1, \quad \mu(\pi i+{\bf B}/2)=a_2, \quad
\mu(\pi i)=a_3.
$$
The function $\mu(z)$ is real along the real axis $\Re u$ and the line
$\ell=\{z=\pi i+z'| z'\in{\Bbb R}\}$, but it is finite only in the second
case.
The function $\mu(\pi i+z')$ changes periodically between $a_2$ and $a_3$
with period ${\bf B}$.
Along the line $\ell$ the differential in (\ref{abel2}) has no zeros,
therefore
$X(\pi i+z')$ and the inverse real function $z'(X)$ are strictly monotonic
functions.
According to (\ref{parX}),
the composition function $U(X)=\mu(\pi i+z'(X))$ is a regular periodic
function
with real period
$$
\frac1{\sqrt{\kappa}}\left[(\rho-c){\bf B}-
\log\frac{\theta_{11}({\bf B}-q/2)}{\theta_{11}({\bf B}+q/2)}\right].
$$
Such a function is shown in figure 1a.
\paragraph{Case 2.} $a_1<a_2<c<a_3$. Now $q,\Pi,{\bf B}$ have the same
expressions as above and
$$
\mu(0)=c, \quad \mu({\bf B}/2)=a_2, \quad \mu(\pi i+{\bf B}/2)=a_1, \quad
\mu(\pi i)=a_3.
$$
The function $\mu(z)$ is real and finite only along the real axis,
on which the differential has a double zero.
As a result, the composition $U(X)=\mu(z(X))$ varies between $a_2$
and $c$, with the same period as above and has cusps for $U=c$.  See
figure 1b.

The cases of other positions of $c$ among the real roots $(a_1,a_2,a_3)$
lead to either one of the above two types of solutions or to unbounded
solutions.

\section{Periodic and solitary peakon solutions.}
\setcounter{equation}{0} In this section we describe different
periodic and solitary solutions to the Dym type and shallow water
SW equations obtained as limits of the traveling wave solutions
and associated with various singularizations of the elliptic
curves described in the previous section. In particular, we
encounter periodic and solitary peakon solutions, i.e. solutions
with discontinuous derivatives.

In contrast to traveling wave solutions given above globally in a
parametric form, peakon solutions can be given {\it explicitly}
only in certain intervals.

\vspace{2mm}

\paragraph{Periodic peakon solution for the Dym type equation.}
Consider again the integral (\ref{abel}). Assuming $\kappa=-1$, in
the limit $a_2\rightarrow c$ the latter can be written in the
equivalent differential form
\begin{equation}
\label{periodpeakon}
\frac{d \lambda}{\sqrt{a_1-\lambda}}=dX, \quad {\rm or} \quad
\frac{d \lambda}{(\lambda-c)\sqrt{a_1-\lambda}}=dX', \quad
dX=(\lambda-c)\, dX',
\end{equation}
$X'$ being a new variable. Now putting $\lambda=a_1-\nu^2$,
$a_1-c=\alpha^2$ and integrating the second differential in
(\ref{periodpeakon}),
we find that
$$
\frac{1}{\alpha}\log\left(\frac{\nu-\alpha}{\nu+\alpha}\right)
=X'+C', \quad C'={\rm const}, \quad {\rm i.e.,} \;\;
\nu=\alpha\,\frac{1+e^{\alpha X'}}{1-e^{\alpha X'}}.
$$
Let us choose here $C'=\pi i/\alpha$. Then we have
$$
\lambda(X' )-c=\alpha_1^2-\alpha_1^2 \left( \frac{
 e^{-\alpha X'/2}- e^{\alpha X'/2} }
{ e^{-\alpha X'/2}+e^{\alpha X'/2} }\right)^2, \quad
{\rm or, equivalently,}
$$
\begin{equation}
\label{Tau}
{\displaystyle
\lambda(X' )-c=4\partial^2_{X'} \log \tau(X' ), \quad
\tau(X' )=e^{-\alpha X'/2}+e^{\alpha X'/2}.}
\end{equation}
This function is $2\pi i$ periodic in $\alpha X$ and
takes real finite values along the real axis.

Next, using the expression (\ref{Tau}), we integrate the relation
$dX=(\lambda(X')-c)\,dX'$ and get the connection between $X$ and
$X'$:
\begin{equation}
\label{variables}
X=4\partial_{X'} \log \tau(X')+X_0 =
2\alpha\frac{e^{-\alpha X'/2}-e^{\alpha X'/2}}
{e^{-\alpha X'/2}+e^{\alpha X'/2}}+X_0, \quad X_0={\rm const}.
\end{equation}
Notice that as $X'\rightarrow -\infty$ or $+\infty$,
$X$ has finite limits which  differ by $4\alpha$.
Thus, expressions (\ref{Tau}) and  (\ref{variables}) give the solution
$U(X)=\lambda(X'(X))$
in a parametric form only in the interval $[X_0-2\alpha,X_0+2\alpha]$.
As follows from  (\ref{Tau}) and  (\ref{periodpeakon}),
at the endpoints of the interval, $U=c$ and ${\rm d}U/{\rm
d}X=\pm\sqrt{a_1-c}$.

Now we define the global solution $U(X)$ for all values of $X$ by
using periodic extension, i.e., by gluing an infinite number of
pieces corresponding to $X_0=4\alpha N$, $N\in\Bbb Z$ in
(\ref{variables}). At the endpoint of each interval the derivative
of the solution changes sign resulting in  a peak, and we obtain a
periodic peakon solution.

On the other hand, a direct integration of the first differential in
(\ref{periodpeakon})
yields the following solution which holds between subsequent peaks
$$
\lambda(X)=a_1-\frac{1}{4} (X-X_0)^2, \quad X\in [X_0-2\alpha, X_0+2\alpha].
$$
Thus, in contrast to $\lambda(X')$ and $X(X')$, the profile of $U(X)$
between peaks
is not exponential, but a quadratic one.

\paragraph{Periodic peakon solution for the SW equation.}
In a similar way, consider a limit of the periodic solution of the SW
equation by putting $a_3=c$ in (\ref{abel2}). Then the curve $\tilde {\cal E}$
becomes singular having a double point at $\mu=c$. Let $\tilde {\cal E}'$
be the corresponding regularized curve.
The integral (\ref{abel2}) gives rise to the following differentials
\begin{equation}
\label{percw}
\frac{d\mu }{\sqrt{(\mu-a_1)(\mu-a_2)}}=dX, \quad
\frac{d\mu }{(\mu -c)\sqrt{(\mu-a_1)(\mu-a_2)}}=dX',
\end{equation}
where, as before, $X=x-ct$, $dX=(\mu-c)dX'$.
Integrating and inverting (\ref{percw}) we obtain respectively
\begin{equation}
\label{direct}
\mu(X)=\frac{a_1-a_2}{4} (e^X+e^{-X})+\frac{a_1+a_2}{2},
\end{equation}
\begin{equation}
\begin{array}{l}
\label{mu-X}
{\displaystyle
\mu(X')-c=\frac 4{(e^{Z/2}+e^{-Z/2})^2/(a_1-c)+(e^{Z/2}-e^{-Z/2})^2/(a_2-c)},}
\\
{\displaystyle
Z=\sqrt{(c-a_1)(c-a_2)}X'.}
\end{array}
\end{equation}
The last expression is $2\pi i$-periodic in $Z$ and provides
a parameterization of the regularized curve $\tilde {\cal E}'$.
Integrating (\ref{mu-X}) with respect to $X'$, we find that
\begin{equation}
\begin{array}{l}
\label{X-X'}
{\displaystyle
X=\log\, \frac{(a_1-a_2)e^{-Z}-(a_1+a_2)+2c-2\sqrt{(c-a_1)(c-a_2)}}
{(a_1-a_2)e^{-Z}-(a_1+a_2)+2c+2\sqrt{(c-a_1)(c-a_2)}} +\rm{const}. }
\end{array}
\end{equation}
Now suppose that $a_1,a_2,c$ are real and $c<a_1<a_2$ or
$a_1<a_2<c$. As follows from (\ref{mu-X}), (\ref{X-X'}), for real
parameter $X'$, the variables $\mu$ and $X$ are real as well. As
$X'\to\pm\infty$, the function $X(X')$ varies between different
finite limits which we denote by $X_1$ and $X_2$, whereas, by
(\ref{mu-X}), $\mu(X')-c$ tends to zero. As a result, the
composition function $U(X)=\mu(X'(X))$ is defined only in the
interval $[X_1,X_2]$, where it takes values in $[c,a_1]$ or
$[a_2,c]$.  According to (\ref{percw}), at the endpoints of the
interval, $U=c$, $dU/dX=\pm\sqrt{(c-a_1)(c-a_2)}$. Like for the
Dym equation, we define the global solution $U(X)$ by periodic
extension of $\mu(X'(X))$. This implies that $U(X)$ has periodic
peaks on the infinite interval $(-\infty, \infty)$.

The expression in (\ref{direct}) provides a piece of the solution between two
subsequent peaks. Solving the equation $\mu(X)=c$ or using (\ref{X-X'})
directly, we find the period of the peakon solution to be
\begin{equation}
\label{period-peak}
X_2-X_1=\log \frac{2c-(a_1+a_2)+2\sqrt{(c-a_1)(c-a_2)}}
{2c-(a_1+a_2)-2\sqrt{(c-a_1)(c-a_2)}}.
\end{equation}
See figure 1c.  This completes the description of the periodic peakon solution.

One can show that the other possible case: $a_1<c<a_2$
does not provide a  real bounded solution. For details about algebraic geometric
approach to describing n-peakon solutions see Alber and Miller [2000] and Alber {\it et al.}
[2000a].

\paragraph{Solitons for the SW equation.} Consider another
possible degeneration of the integral (\ref{abel2}), assuming that
$a_2=a_3=a$ and
$a,a_1,c$ are distinct. Then we have
\begin{equation}
\label{solcw}
\begin{array}{l}
{\displaystyle
\frac{(\mu-c)\,d\mu }{(\mu-a)\sqrt{(\mu-a_1)(\mu-c)}}=dX, \quad
\frac{d\mu }{(\mu-a)\sqrt{(\mu-a_1)(\mu-c)} }=dX',} \\
\\
{\displaystyle
X=x-ct+{\rm const}, \quad dX=(\mu-c)dX'.}
\end{array}
\end{equation}
The first differential has 2 pairs of simple poles on the rational curve \\
$\{\nu^2=(\mu-a_1)(\mu-c)\}$ corresponding to $\mu=a$ and $\mu=\infty$ and
a double zero for $\mu=c$.
Therefore the inverse function $U(X)$ again has branching of the form
(\ref{branching}).

Integrating the second differential in (\ref{solcw}) and inverting, we obtain
\begin{equation}
\label{mu-tildeZ}
\mu(X')-a=\frac 4{(e^{{\tilde Z}/2}+e^{-{\tilde Z}/2})^2/(c-a)
+(e^{{\tilde Z}/2}-e^{-{\tilde Z}/2})^2/(a_1-a)},
\end{equation}
$$
{\tilde Z}=\sqrt{(a-a_1)(a-c)}X'.
$$
Then, integrating the relation between $X$ and $X'$ yields
\begin{equation}
\label{X-X''}
\begin{array}{l}
{\displaystyle
X=(a-c)X'+\log\,\frac{(a_1-c)e^{-{\tilde Z}}-(a_1+c)+2a-2\sqrt{(a-a_1)(a-c)}}
{(a_1-c)e^{-{\tilde Z}}-(a_1+c)+2a+2\sqrt{(a-a_1)(a-c)}} }
{\displaystyle
+\rm{const}. }
\end{array}
\end{equation}

As before, let us consider possible real solutions assuming that $a,a_1,c$ are
real. According to (\ref{mu-tildeZ}), $\mu(\tilde Z)$ is $2\pi i$-periodic
and it takes real values along the lines Im${\tilde Z}=N\pi i$, $N\in\Bbb Z$.

\paragraph{Case 1.} $a<a_1<c$. The function $\mu(\tilde Z)$ is real and finite
only along the lines Im${\tilde Z}=\pi i+2N\pi i$, where it
describes a smooth solitary wave tending to
$a$ as $\Re X$ or $\Re{\tilde Z}$ tends to $\pm\infty$ and having the
maximum $a_1$.
Along these lines the derivative $dX/dX'=\mu-c$ is always real, negative and
separated from zero.
Therefore, the composition function $U(X)=\mu(X'(X))$ gives
a smooth solitary wave as well.

\paragraph{Case 2.} $a<c<a_1$. Now $\mu(\tilde Z)$ is real and finite
only along the lines Im${\tilde Z}=2N\pi i$, where it again
describes a smooth solitary wave tending to $a$ as $\Re X\to
\pm\infty$ and having the maximum $c$. On the other hand, along
these lines the derivative $dX/dX'=\mu-c$ is always real and
negative except when $\mu=c$. At this point $dX/dX'$ has a double
zero. As a result, for $\mu=c$ the derivative $dU/dX$ blows up and
the composition function $U(X)=\mu(X'(X))$ describes a solitary
cusp (cuspon).

\paragraph{Solitary peakon for the SW equation.}
As seen from (\ref{period-peak}), in the limit $a_1=a_2=a$ the
period of the above peakon solution tends to infinity. This gives
us a solitary peakon solution, which can also be regarded as the
separatrix between the smooth soliton and cuspon solutions.
Indeed, in this case the algebraic curve $\tilde{\cal E}$ has 2
double points with $\mu=a$ and $\mu=c$. Its reauthorization
$\widehat{\cal E}$ consists of two disjoint copies of ${\Bbb
P}=\{\mu\}\cup\infty$. On them the differentials (\ref{percw})
take the following simple form
\begin{equation}
\label{diff-peak}
\frac{d\mu}{\mu-a}=l\,dX, \quad \frac{d\mu}{(\mu-c)(\mu-a)}=l\,dX',
\end{equation}
where, as above, $dX=(\mu-c)dX'$ and $l=\pm 1$, regarding to the copy of
$\Bbb P$.
First, we suppose $l=1$. After integration in $\mu$ and inversion this yields
\begin{equation}
\label{peak-X}
\mu(X)=a+e^{X+C_2}, \quad C_2={\rm const}, \qquad
\mu(X')=\frac{a-ce^{(a-c)X'}}{1-e^{(a-c)X'}} .
\end{equation}
The function $\mu(X')$ has period $2\pi i/(a-c)$ and its fundamental
domain gives a parameterization of the cylinder
${\Bbb P}\setminus\{\mu=a\}\setminus\{\mu=c\}$.

Integrating the differential relation between $X$ and $X'$, we
obtain $$ X(X')=-\int_{X_0'}^{X'} \frac{(a-c)
e^{-(a-c)X'}}{1-e^{-(a-c)X'}}\,dX' =-\log (e^{(c-a)X'}-1)+C_1,
\quad X_0', C_1={\rm const}. $$

Suppose that $a,c$ are real and, for definiteness, put $a<c$. The
function $\mu(X')$ is $2\pi i/(a-c)$ periodic, it has a pole along
the real axis, whereas along the line Im$X'=\pi i/(c-a)$ it varies
in the interval $(a,c)$. Let us put $C_1=\pi i$. Then $X(X')$ is
real along this line. We notice that as $\Re X'\rightarrow
\infty$, $X$ tends to $-\infty$ and $\mu$ tends to $a$, whereas
for $\Re X'\rightarrow -\infty$, we have $X \rightarrow 0$ and
$\mu \rightarrow c$. Thus the composition function $\mu(X'(X))$ is
defined on the interval $[-\infty,0]$ only, where it is also given
by the first expression in (\ref{peak-X}) for $C_2=\log(c-a)$.

Now assuming $l=-1$, we obtain the same expressions with $X$
replaced by $-X$. Let us choose the same values of the integration
constants $C_1,C_2$. Then the composition function $\mu(X'(X))$ is
defined on the interval $[0,\infty]$, where it has the form
$\mu(X)=a+(c-a)e^{-X}$. As a result, the two copies of ${\Bbb P}$
give rise to two branches of the real continuous solution $$
U(X)=a+(c-a)e^{-|X|}, \qquad X\in\Bbb R, $$ which has a peak at
the origin.
\paragraph{Remark.} As seen from the above peakon solutions, the amplitude
of peaks coincides with the velocity $c$, which is a specific
property of soliton propagation.

\section{Stationary quasi-periodic solutions.}
\setcounter{equation}{0} Stationary solutions provide profiles of
the quasi-periodic wave solutions. For the sake of clarity, in
this paper we restrict ourselves to the simplest nontrivial case
$g=2$. All the formulae and solutions below (except structure of
real solutions) can be easily extended to the arbitrary
$g$-dimensional case which is only notationally more complicated.

\paragraph{Stationary quasi-periodic solutions for the SW equation.}
According to the trace formula, in the genus $2$ case we have
\begin{equation}
\label{trace0}
U(x,t)=\mu_1+\mu_2 - \sum_{j=1}^5 a_j,
\end{equation}
and equations (\ref{A-J-g}) take the form
\begin{equation}
\label{A-J}
\begin{array}{l}
{\displaystyle
\frac{\mu_1 d\mu_{1}} {2\sqrt{R_6(\mu_1)} }
+\frac{\mu_2d\mu_{2}} {2\sqrt{R_6(\mu_2)} }=dt, } \\%
\\
{\displaystyle
\frac{\mu_1^2 d\mu_{1}}{2\sqrt{R_6(\mu_1)}}
+\frac{\mu_{2}^2d\mu_{2}} { 2\sqrt{R_6(\mu_2)}}=dx,}
\end{array}
\end{equation}
where
$$
R_6(\mu)=-\kappa\mu(\mu-a_1)\cdots(\mu-a_5), \quad a_1,\dots,a_5={\rm const}.
$$
Here we suppose that all the roots of $R_6(\mu)$ are distinct.
The variables $\mu_{1},\mu_{2}$ must be regarded as coordinates of points
$P_{1}=(\mu_{1},w_{1})$, $P_{2}=(\mu_{2},w_{2})$ on the genus 2 hyperelliptic
curve $\Gamma=\{w^2=R_6(\mu)\}$. Equations (\ref{A-J}) involve one holomorphic
differential and one meromorphic differential of the third kind having a pair
of simple poles at the infinite points $\infty_-,\infty_+$ on $\Gamma$.
Integrating (\ref{A-J}), we obtain the mapping of the symmetric product
$\Gamma^{(2)}$ to ${\Bbb C^2}=(t,x)$
\begin{equation}
\label{x-t-map}
\begin{array}{l}
{\displaystyle
\int_{P_0}^{P_1} \frac{\mu\,d\mu}{ 2\sqrt{R_6(\mu)}}
+\int_{P_0}^{P_2}\frac{\mu\,d\mu}{ 2\sqrt{R_6(\mu)}}=t,} \\%
\\
{\displaystyle
\int_{P_0}^{P_1}\frac{\mu^2 d\mu}{2\sqrt{R_6(\mu)}}
+\int_{P_0}^{P_2}\frac{\mu^2 d\mu}{ 2\sqrt{R_6(\mu)}}=x. }
\end{array}
\end{equation}
where $P_0$ is a fixed basepoint of the mapping. Notice that for
$P_1$ or $P_2=\infty_-,\infty_+$, we have $x=\infty$.
Let us fix a canonical basis of cycles $A_1,A_2,B_1,B_2$ on $\Gamma$ in
a standard way (see, for example, Mumford [1978]).
The mapping has four independent (over the reals) 2-dimensional vectors
of periods of the above differentials along the cycles. In addition, it
has one extra period vector corresponding to a homology zero cycle around
$\infty_-$ or $\infty_+$. As a result, the mapping (\ref{x-t-map})
has 5 period vectors in ${\Bbb C}^2$ hence its inversion is not well defined:
there do not exist meromorphic
functions on ${\Bbb C}^2$ with five periods. In particular,
$U(x,t)$ is not a meromorphic or single valued complex function of $t,x$.

In order to describe properties of $U(x,t)$, we,
first, fix time by putting $t=t_0$ ($dt=0$) and consider stationary solutions
$U(x,t_0)$. Introduce a new coordinate $x'$ such that
\begin{equation}
\label{change}
dx=\mu_1\mu_2\,dx'.
\end{equation}
Then equations (\ref{A-J}) lead to the Abel--Jacobi mapping
of $\Gamma^{(2)}$ to the Jacobian variety Jac($\Gamma$)
of $\Gamma$, which includes holomorphic differentials only:
\begin{equation}
\label{A-J1}
\begin{array}{l}
{\displaystyle
\int_{P_0}^{P_1} \frac{d\mu}{ 2\sqrt{R_6(\mu)}}
+\int_{P_0}^{P_2}\frac{d\mu}{ 2\sqrt{R_6(\mu)}}=u_1,} \\%
\\
{\displaystyle
\int_{P_0}^{P_1}\frac{\mu d\mu}{2\sqrt{R_6(\mu)}}
+\int_{P_0}^{P_2}\frac{\mu d\mu}{ 2\sqrt{R_6(\mu)}}=u_2, } \\%
\\
{\displaystyle
u_1=x'+{\rm const}, \quad u_2={\rm const}, }
\end{array}
\end{equation}
where $u_1, u_2$ are coordinates on the universal covering
${\Bbb C}^2$ of Jac$(\Gamma)$.

Let $\bar{\omega}_1, \bar{\omega}_2$ be the dual basis of {\it normalized}
holomorphic differentials on $\Gamma$ with respect to the above choice
of cycles and $z_1,z_2$ be the corresponding
coordinates on the universal covering of Jac$(\Gamma)$:
\begin{equation}
\label{norm}
\begin{array}{l}
{\displaystyle
\bar{\omega}_1=\frac{d_{11}+d_{12}\mu}{2\sqrt{R_6(\mu)} }\, d\mu, \quad
\bar{\omega}_2=\frac{d_{21}+d_{22}\mu}{ 2\sqrt{R_6(\mu)}}\, d\mu, } \\%
\\
{\displaystyle
z_1=d_{11}u_1+d_{12}u_2, \quad z_2=d_{21}u_1+d_{22}u_2.}
\end{array}
\end{equation}
Here the normalizing constants $d$ are uniquely determined by the conditions
$\oint_{A_i}\bar{\omega_j}=\delta_{ij}$.

Recall that the standard theta-function related to a Riemann surface
of genus $g$ and theta-functions with
characteristics $\alpha=(\alpha_{1},\dots,\alpha_{g})$,
$\beta=(\beta_{1},\dots,\beta_{g})\in{\Bbb R}^{g}$
have the form
\begin{equation}
\label{theta}
\begin{array}{l}
{\displaystyle
\theta(z|{\bf B})=\sum_{M\in{\Bbb Z}^{g} }\exp \left(\frac12({\bf
B}M,M)+(M,z)\right), }\\
{\displaystyle
(M,z)=\sum^{g}_{i=1} M_{i}z_{i}, \quad
({\bf B}M,M)=\sum^{g}_{i,j=1}{\bf B}_{ij}M_{i}M_{j}, } \\
\\
{\displaystyle
\theta\left[\begin{array}{c} \alpha\\ \beta\end{array}\right](z|{\bf B})
=\exp \{ ({\bf B}\alpha,\alpha)/2+(z+2\pi i\beta,\alpha) \}
\theta(z+2\pi i\beta+{\bf B}\alpha|{\bf B}), } \end{array}
\end{equation}
${\bf B}$ being the $g\times g$ period matrix of $\Gamma$. In the sequel we
shall
omit it in the notation.

Now we choose the basepoint $P_0$ of the mapping (\ref{A-J1}) to be the
last Weierstrass point $(a_5,0)$ on $\Gamma$. Then,
according to the trace formula for even order hyperelliptic curves
(see e.g., Clebsch and Gordan [1866], Dubrovin [1981])
\begin{equation}
\label{trace}
U=\mu_1+\mu_2 - \sum_{j=1}^5 a_j={\rm const}-\partial_W
\log\frac{\theta[\delta](z-q/2)}
{\theta[\delta](z+q/2)},
\end{equation}
$$ z=(z_{1},z_{2}), \quad q=(q_{1},q_{2})^{T}, \quad
q_{i}=\int^{\infty_+}_{\infty_-}\bar{\omega}_{i}, $$ where, in
view of the normalizing change (\ref{norm}), $z_1=d_{11}x'+{\rm
const}$, $z_2=d_{21}x'+{\rm const}$, $\partial_W$ is the
derivative along a tangent vector $W$ of
$\Gamma\subset\rm{Jac}(\Gamma)$ at $\infty_+$, namely, in the
coordinates $(u_1,u_2)$, $W=(0,1)^T$, and in the coordinates
$(z_1,z_2)$, $W=(d_{12},d_{22})^T$. Finally,
$\delta=(\delta'',\delta')^T$, $\delta'',\delta'\in\frac 12{\Bbb
Z}^{g}/{\Bbb Z}^{g}$ is the half-integer theta-characteristic
corresponding to the vector of Riemann constants (see Mumford
[1983]). For the chosen standard canonical basis of cycles and the
basepoint $P_0=(a_5,0)$
\begin{equation}
\label{Delta}
\delta '=(1/2,\dots,1/2)^{T},\quad \delta''=(g/2,(g-1)/2,\dots,1,1/2)^{T}
\;\, (\rm{mod}\; 1).
\end{equation}
Thus, in our case
$$ \delta=\left( \begin{array}{cc} 0  & 1/2 \\
                                  1/2 & 1/2 \end{array} \right). $$
The function $U(z_1,z_2)$ is meromorphic on Jac$(\Gamma)$ and it has simple
poles along 2 translates of the theta-divisor
$\Theta=\{\theta(z)=0\}\subset\rm{Jac}(\Gamma)$:
$$
\Theta_-=\{\theta[\delta](z-q/2)=0\}, \quad
\Theta_+=\{\theta[\delta](z+q/2)=0\},
$$
which are tangent to each other at the origin $\{z=0\}$.
Thus, $U(z_1(x'),z_2(x'))$ is a quasi-periodic function of the
complex variable $x'$. Notice that a quasi-periodic genus 2 solution of the
nonlinear mKdV equation has the same form.

We also notice that the point $E_0=(\mu=0,w=0)$
is a Weierstrass (branch) point on $\Gamma$. Then, following Clebsch and
Gordan
[1866], we have the following expression for the symmetric polynomial
\begin{equation}
\label{mu}
\mu_1\mu_2=\varrho \frac{\theta^2[\delta+\eta_0](z)}
{\theta[\delta](z+q/2)\theta[\delta](z-q/2)}, \quad \varrho={\rm const},
\end{equation}
where $\eta_0$ is the half-integer theta-characteristic corresponding to
the branch point $E_0$:
\begin{equation}
\label{eta0}
\eta_0=(\eta_0'',\eta_0')^T, \quad
\int_{P_0}^{E_0}(\bar{\omega}_1,\bar{\omega}_2)^T
=2\pi i\, \eta_0''+{\bf B}\eta_0'\in {\Bbb C}^2.
\end{equation}

Thus, the product $\mu_1\mu_2$ is a meromorphic
function on Jac$(\Gamma)$ having simple poles along $\Theta_-,\Theta_+$
and a {\it double} zero along another translate of the theta-divisor $\Theta$,
$\Theta_0=\{\theta[\delta+\eta_0](z)=0\}$, passing through the origin and
intersecting each of the translates $\Theta_-,\Theta_+$ at two points.
The translate $\Theta_0$ can also be interpreted as the
image of the curve $\Gamma$ itself under the Abel--Jacobi mapping
(\ref{A-J1}):
$$
\Theta_0=\left\{\int_{P_0}^P(\bar{\omega}_1,\bar{\omega}_2)^T
+\int_{P_0}^{E_0}(\bar{\omega}_1,\bar{\omega}_2)^T\bigg | P\in\Gamma\right\}.
$$
It follows from (\ref{change}), (\ref{mu}) that generically the derivative
 of the function $x(x')$ is equal to $\mu_1\mu_2$ and that it has  a double
zero each
time when the complex $x'$-flow
intersects $\Theta_0$, i.e., when $\theta[\delta+\eta_0](z)$ vanishes,
except the points where the flow is tangent to $\Theta_0$, i.e., when
$\theta[\delta+\eta_0](z)$ has a higher vanishing order in $x'$ and
$\mu_1\mu_2$ too. This takes place only at the origin of Jac$(\Gamma)$.
Since at the origin the solution (\ref{trace}) blows up,
we conclude that for bounded solutions the function $\mu_1\mu_2$
may have only a double zero and $x(x')$ a simple zero in $x'$.

On the other hand, in view of the second equation of (\ref{x-t-map}),
the original variable $x$ is a sum of Abelian integrals of third kind.
Introduce the normalized differentials of third kind
$\Omega_{\infty_{-}\infty_{+}}$ on $\Gamma$ having poles at
$\infty_{-},\infty_{+}$ with residues $\pm 1$:
\begin{equation}
\label{Omega}
\Omega_{\infty_{-}\infty_{+}}
=\frac{\mu^2\, d\mu}{\sqrt{R_6(\mu)}}+h_1\bar{\omega}_1+h_2\bar{\omega}_2,
\end{equation}
where $h_1,h_2$ are normalizing constants specified by the conditions for
$\Omega_{\infty_{-}\infty_{+}}$ to have zero $A$-periods on $\Gamma$.

According to Clebsch and Gordan [1866],
\begin{equation}
\label{AB-trans-3}
\int_{P_0}^{P_1}\Omega_{\infty_{-}\infty_{+}}+
\int_{P_0}^{P_2}\Omega_{\infty_{-}\infty_{+}}
=\log\frac{\theta[\delta](z+q/2)}{\theta[\delta](z-q/2)}+{\rm const}.
\end{equation}
Then, in view of the second equation in (\ref{A-J}) and (\ref{Omega}), we get
\begin{equation}
\label{trans}
\begin{array}{c}
{\displaystyle
x(x')=\log\frac{\theta[\delta](z+q/2)}{\theta[\delta](z-q/2)}-h_1z_1
-h_2z_2+{\rm const}, } \\%
\\
{\displaystyle
z_1=d_{11}x'+{\rm const}, \quad z_2=d_{21}x'+{\rm const}. }
\end{array}
\end{equation}

As a result, we expressed the stationary quasi-periodic solution
$U$ and the argument $x$ in terms of the auxiliary complex
variable $x'$. The algebraic geometrical structure of the general
solution $U(x,t)$ and the behaviour of real solutions will be
considered in the next sections.
\medskip

\paragraph{Stationary quasi-periodic solutions for the Dym equation.}

Now we pass to the Dym equation (\ref{dym-sw-eqn}) and seek its solutions
again in the form (\ref{trace0}). In this case
the variables $\mu_1,\mu_2$ again change according to equations of the form
(\ref{A-J}) with the only difference being that the order of the polynomial
defining the corresponding hyperelliptic curve is odd:
\begin{equation}
\label{A-J-odd}
\begin{array}{l}
{\displaystyle
\frac{\mu_1 d\mu_{1}}{2\sqrt{R_5(\mu_1)} }
+\frac{\mu_2d\mu_{2}}{2\sqrt{R_5(\mu_2)}}=dt, } \\%
\\
{\displaystyle
\frac{\mu_1^2 d\mu_{1}} {
2\sqrt{R_5(\mu_1)}}+\frac{\mu_{2}^2d\mu_{2}} { 2\sqrt{R_5(\mu_2)}}=dx, } \\%
\\
{\displaystyle
R_5(\mu)=-\kappa\mu(\mu-a_1)\cdots(\mu-a_4), }
\end{array}
\end{equation}
hence the corresponding hyperelliptic
curve $\Gamma=\{w^2=R_5(\mu)\}$ has just one infinite point $\infty$. As
a consequence, the equations (\ref{A-J-odd}) contain one holomorphic
differential and one differential of the {\it second} kind.

As before, we first consider stationary solutions by putting
$t=t_0$ $(dt=0)$ and assuming $\kappa=1$. Notice that under these
conditions, (\ref{A-J-odd}) has the same structure as quadratures
for the Jacobi problem on geodesics on a triaxial ellipsoid $Q$,
where $\mu_1,\mu_2$ play the role of ellipsoidal coordinates on
$Q$, parameters $a_1,a_2,a_3$ the squares of the semi-axes of $Q$,
$a_4$ the constant of motion, and $x$ the length of a geodesic.

Under the change of parameter (\ref{change}), we arrive at the Abel--Jacobi
mapping
\begin{equation}
\label{A-J4}
\begin{array}{l}
{\displaystyle
\int_{P_0}^{P_1} \frac{d\mu}{ 2\sqrt{R_5(\mu)}}
+\int_{P_0}^{P_2}\frac{d\mu}{ 2\sqrt{R_5(\mu)}}=u_1,} \\%
\\
{\displaystyle
\int_{P_0}^{P_1}\frac{\mu d\mu}{2\sqrt{R_5(\mu)}}
+\int_{P_0}^{P_2}\frac{\mu d\mu}{ 2\sqrt{R_5(\mu)}}=u_2, } \\%
\\
{\displaystyle
u_1=x'+{\rm const}, \quad u_2={\rm const}, }
\end{array}
\end{equation}
This change was first made by Weierstrass [1878]
in order to find the theta-functional solution for the geodesic problem
(see also Cewen [1990]). Next we introduce normalized holomorphic
differentials $\bar{\omega}_1,\bar{\omega}_2$ on $\Gamma$
and coordinates $z_1,z_2$ on the universal covering of
Jac$(\Gamma)$ according to (\ref{norm}) and, in addition,
the normalized differential of the
second kind having a double pole at $\infty$
\begin{equation}
\label{Omega2}
\Omega_{\infty}^{(1)}=\frac{\mu_i^2 d\mu_{i}}{2\sqrt{R_5(\mu_i)}}
+h_1'\bar{\omega}_1+h_2'\bar{\omega}_2.
\end{equation}
Similarly to (\ref{Omega}), the constants $h_1',h_2'$ are uniquely defined
by the condition that  $\Omega_{\infty}^{(1)}$  have zero $A$-periods on
$\Gamma$.

Then, instead of the expressions (\ref{trace}), and (\ref{mu}), we
have (see, e.g., Dubrovin [1981], Dubrovin {\it et al.} [1985])
\begin{equation}
\label{trace2}
\begin{array}{l}
{\displaystyle
U(x')=\mu_1+\mu_2={\rm const}-\partial_{V}^2 \theta[\delta](z), } \\%
\\
{\displaystyle
z_1=d_{11}x'+{\rm const}, \quad z_2=d_{21}x'+{\rm const}, }
\end{array}
\end{equation}
where $\partial_V$ is the derivative along the tangent vector $V$ of
$\Gamma\in$Jac$(\Gamma)$ at $\infty$: $V=(d_{12},d_{22})^T$, and,
respectively,
\begin{equation}
\label{mu1}
\mu_1\mu_2=\kappa\,\frac{\theta^2[\delta+\eta_0](z)}{\theta^2[\delta](z)},
\qquad \kappa={\rm const},
\end{equation}
where the characteristic $\eta_0$ is defined in (\ref{eta0}).
In addition, in contrast to (\ref{AB-trans-3}), the sum of Abelian integrals
of second kind has the form
\begin{equation}
\label{AB-trans-2}
\int_{P_0}^{P_1}\Omega_{\infty}^{(1)}+
\int_{P_0}^{P_2}\Omega_{\infty}^{(1)}={\rm const}
-\partial_{V}\log\theta[\delta](z).
\end{equation}
Comparing this with (\ref{Omega2}), we find that
the analog of the relation (\ref{trans}) between the parameters $x$ and $x'$
has the form
\begin{equation}
\label{trans1}
\begin{array}{l}
{\displaystyle
x(x')=-\partial_{V}\log\theta[\delta](z)-h_1z_1-h_2z_2+{\rm const}, } \\%
\\
{\displaystyle
z_1=d_{11}x'+{\rm const}, \quad z_2=d_{21}x'+{\rm const}. }
\end{array}
\end{equation}
This expression can be regarded as a 2-dimensional generalization of
the Weierstrass zeta-function in (\ref{z1}).

Thus, we have expressed the stationary solution $U$ and the
argument $x$ in terms of the auxiliary complex variable $x'$.
Various types of real solutions defined by the above expressions
will be considered in Section 6.

\section{Time-dependent quasi-periodic solutions.}
\setcounter{equation}{0}

\paragraph{The solutions for the SW equation.}

In order to obtain general time-dependent solutions $U(x,t)$ of
the SW equation
given by (\ref{trace0}), one has to invert the mapping (\ref{x-t-map}).
However, as already mentioned, the problem of inversion is unsolvable in
terms of meromorphic functions.

To describe the structure of general solutions,
let us first consider a divisor of 3 points $P_i=(\mu_i,w_i)$, $i=1,2,3$ on
$\Gamma\setminus\{\infty_-,\infty_+\}$ and the following extended equations
\begin{equation}
\label{A-J2}
\sum_{i=1}^3\frac{d\mu_i}{2\sqrt{R_6(\mu_i)}}=dy, \quad
\sum_{i=1}^3\frac{\mu_{i}d\mu_{i}}{2\sqrt{R_6(\mu_1)}}=dt, \quad
\sum_{i=1}^3\frac{\mu_i^2 d\mu_{i}}{2\sqrt{R_6(\mu_i)}}=dx,
\end{equation}
including the extra variable $y$, two holomorphic differentials and one
differential of the third kind on $\Gamma$. The latter are linear
combinations of the normalized differentials
$\bar{\omega}_1, \bar{\omega}_2, \Omega_{\pm\infty}$ defined in (\ref{norm})
and (\ref{Omega}). According to Clebsch and Gordan [1866], Fedorov [1999],
equations (\ref{A-J2}) describe a differential of a well defined mapping of
the symmetric product $(\Gamma\setminus\{\infty_-,\infty_+\})^{(3)}$ to
{\it generalized Jacobian variety} Jac$(\Gamma,\infty_{\pm})$, a
{\it noncompact} algebraic group represented as the quotient of ${\Bbb C}^3$
by a lattice $\Lambda$ generated by five vectors of periods of the
differentials
$\bar{\omega}_1, \bar{\omega}_2, \Omega_{\pm\infty}$ on $\Gamma$.
Topologically, Jac$(\Gamma,\infty_{\pm})$ is the product of the
2 dimensional variety Jac$(\Gamma)$ and the cylinder
${\Bbb C}^{*}={\Bbb C}\setminus\{0\}$. An analytical and algebraic-geometrical
description of generalized Jacobians can be found in Clebsch and Gordan
[1866],
Belokolos et all [1994], Fedorov [1999], Gavrilov [1999].

Let $z_1,z_2,Z$ be coordinates on the universal covering of
Jac$(\Gamma,\infty_{\pm})$ such that
\begin{equation}
\label{AJG}
\sum_{i=1}^3\int_{P_0}^{P_i} \bar{\omega}_1=z_1, \quad
\sum_{i=1}^3\int_{P_0}^{P_i} \bar{\omega}_2=z_2, \quad
\sum_{i=1}^3\int_{P_0}^{P_i}\Omega_{\pm\infty}=Z,
\end{equation}
where, as above, $P_0=(a_5,0)$. Then, according to (\ref{norm}),
(\ref{Omega}),
\begin{equation}
\label{norm1}
\begin{array}{l}
{\displaystyle
z_1=d_{11}y+d_{12}t+{\rm const}, \quad z_2=d_{21}y+d_{22}t+{\rm const}, } \\%
{\displaystyle
Z=x+h_1(d_{11}y+d_{12}t)+h_2(d_{21}y+d_{22}t)+{\rm const}. }
\end{array}
\end{equation}

The problem of inversion of Abel--Jacobi mappings including differentials
of the third and second kind is solved in terms of {\it generalized
theta-functions} which are finite sums of products of customary
theta-functions, rational functions, and exponentials
(see Ercolani [1989], Fedorov [1999], Gagnon {\it et al.} [1987]).
To invert the mapping (\ref{AJG}), we shall make use of the following
theta-functions
\begin{equation}
\label{gen-theta}
\begin{array}{l}
{\displaystyle
\widetilde{\theta}(z,Z)=e^{Z/2}\theta(z+q/2)+e^{-Z/2}\theta(z-q/2), } \\%
\\
{\displaystyle
\widetilde{\theta}[\eta](z,Z)=e^{Z/2}\theta[\eta](z+q/2)
+e^{-Z/2}\theta[\eta](z-q/2), }
\end{array}
\end{equation}
$$
z=(z_1,z_2), \quad q=(q_{1},q_{2})^{T}, \quad
q_{1}=\int^{\infty_+}_{\infty_-}\bar{\omega}_{1}, \quad
q_{2}=\int^{\infty_+}_{\infty_-}\bar{\omega}_{2},
$$
where $\theta(z),\theta[\eta](z)$ are customary theta-functions associated
with the curve $\Gamma$ with half-integer theta-characteristics $\eta$.
Like $\theta[\eta](z)$, generalized theta-functions have a quasi-periodic
property: a shift of the argument $(z,Z)$ by any period vector of
the generalized Jacobian results in multiplication of
$\widetilde{\theta}[\eta](z,Z)$ by a constant factor.

Now consider the dissection $\widetilde\Gamma$ of $\Gamma$ along
the canonical cycles $A_{1},A_2,B_1,B_{2}$, which is a
one-connected domain having the form of an octagon. In addition,
we cut $\widetilde{\Gamma}$ along the paths joining a point $O$ on
the boundary $\partial\widetilde{\Gamma}$ of $\widetilde{\Gamma}$
to the points $\infty_{-},\infty_{+}$. On the obtained domain
$\widetilde{\Gamma}'$ we introduce the single-valued function
$\widetilde{F}(P)=\widetilde{\theta}[\delta](\widetilde{{\cal
A}}(P)-(z,Z)^T)$, where $$ \widetilde{{\cal A}}(P)=\left(
\int_{P_0}^{P}\bar{\omega}_1,\int_{P_0}^{P}\bar{\omega}_2,
\int_{P_0}^{P}\Omega_{\infty_\pm}\right)^T, $$ and the
characteristic $\delta$ is defined in (\ref{Delta}). Then the
following analog of the Riemann theorem holds (see e.g., Fedorov
[1999], Gagnon {\it et al.} [1985]).

\begin{thm} \label{Riemann}
Let the coordinates $z,Z$ be such that the function
$\widetilde{F}(P)$ does not vanish identically on $\widetilde{\Gamma}'$.
Then it has precisely 3 zeros $P_{1},P_{2},P_{3}$ giving a unique solution
to the inversion of the generalized mapping (\ref{AJG}).
\end{thm}
Now let us consider the logarithmic differential
$\mu(P)d\log\widetilde{F}(P)$. By Theorem \ref{Riemann}, the sum
of the residues of its poles in the domain $\widetilde{\Gamma}'$
equals $\mu(P_{1})+\mu(P_2)+\mu(P_{3})$. Applying the residue
theorem, after calculations, we get the following compact ``trace
formula''
\begin{equation}
\label{3trace}
\mu_1+\mu_2+\mu_3={\rm const}
-\frac{e^Z \theta[\delta](z+q)+e^{-Z}\theta[\delta](z-q)}{\theta[\delta](z)}
\end{equation}
with the characteristic $\delta$ specified in (\ref{Delta}).

The principal difference between the extended mappings (\ref{A-J2}) or
(\ref{AJG}) and the system (\ref{x-t-map})
is that the latter contains only 2 points on
$\Gamma\setminus\{\infty_-,\infty_+\}$. On the other hand, (\ref{A-J2})
reduces
to (\ref{A-J}) by fixing $P_3\equiv P_0$ ($\mu_3\equiv a_5$, $d\mu_3\equiv
0$).
Under this condition, (\ref{AJG}) describes the embedding of the symmetric
product
$(\Gamma\setminus\{\infty_-,\infty_+\})^{(2)}$ into
Jac$(\Gamma,\infty_{\pm})$.
Its image is a 2-dimensional {\it nonlinear} analytic subvariety (stratum)
$W_2$.
Like the generalized Jacobian itself, it is a noncompact variety.
\medskip

\paragraph{Remark 5.1.} In case of customary Jacobian varieties,
the corresponding nonlinear subvarieties and their stratification
have been studied in Gunning [1972] and Vanhaecke [1995].
Such varieties or their open subsets often appear as (coverings of)
complex invariant
manifolds of finite-dimensional integrable systems (see Vanhaecke [1995],
Abenda and Fedorov [1999]).

It follows from the above that on the stratum $W_2$ the variables
$z_1,z_2,Z$ play the role of excessive (abundant)  coordinates, hence
they cannot be independent there.
The
analytic structure of $W_2$ is explicitly described by the following theorem
(see e.g., Fedorov [1999], Gagnon {\it et al.} [1987]).

\begin{thm} \label{zero-locus}
The subvariety $W_2\subset{\rm Jac}(\Gamma,\infty_{\pm})$ coincides
with the zero locus of the generalized theta-function:
\begin{equation}
\label{W2}
W_2=\{e^{Z/2}\theta[\delta](z+q/2)-e^{-Z/2}\theta[\delta](z-q/2)=0\}.
\end{equation}
\end{thm}

On the other hand, in view of relations (\ref{norm1}), the coordinates $z,Z$
are linear functions of the variables $x,t$, and $y$.
Thus, equation (\ref{W2}) can also be regarded as a constraint on them.
It follows that on fixing $P_3=P_0$, $y$ becomes a transcendent function of
$x,t$.

Now we notice that the sum $\mu_1+\mu_2+a_5=\mu(P_1)+\mu(P_2)+\mu(P_0)$
coincides with the restriction of the total sum $\mu(P_1)+\mu(P_2)+\mu(P_3)$,
as a function on Jac$(\Gamma,\infty_{\pm})$, onto $W_2$. Then, using
expression (\ref{3trace}), we conclude that the 2-phase solution of the
SW equation has the form
\begin{equation}
\label{trace-SW}
U(x,t)={\rm const}
-\frac{e^Z \theta[\delta](z+q)+e^{-Z}\theta[\delta](z-q)}{\theta[\delta](z)}
\end{equation}
$$ z_1=d_{11}y+d_{12}t, \quad z_2=d_{21}y+d_{22}t, \quad
Z=x+h_1(d_{11}y+d_{12}t)+h_2(d_{21}y+d_{22}t),
$$
where the extra variable $y$ depends on $x,t$ according to (\ref{W2}).
As a result, we arrive at the following algebro-geometric description of
motion:
\vspace{3mm}

\noindent
{\it The $x$-flow ($t$-flow) defined by equations (\ref{A-J})
evolves on the nonlinear variety ${W}_2\subset{\rm Jac}(\Gamma,\infty_{\pm})$
in such a way that $y$ is a nonlinear transcendent function of $x$
(respectively, of $t$). In this sense the flow is nonlinear.}
\vspace{3mm}

\noindent
We emphasize that the solution $U(x,t)$ is neither meromorphic in $x$,
nor in $t$.

\paragraph{Remark 5.2.} Let us consider the $x$-flow by putting $t=$const.
It turns out that, up to an additive constant, the extra variable
$y$ can now be identified with the auxiliary variable $x'$
introduced in (\ref{change}) when we considered stationary
solutions. Indeed, in view of (\ref{norm1}), in this case the
condition in (\ref{W2}) becomes
\begin{equation}
\label{Zz}
Z=x+h_1z_1+h_2z_2+{\rm const}
=\log\frac{\theta[\delta](z-q/2)}{\theta[\delta](z+q/2)}+{\rm const},
\end{equation}
which is equivalent to the relation (\ref{trans}) between $x$ and $x'$.
In view of (\ref{Zz}) and the addition
theorem for theta-functions, the solution (\ref{trace-SW})
reduces to the stationary solution (\ref{trace}).

In contrast to $x$, the parameter $t$ enters both expressions for $z$
and $Z$ in
(\ref{trace-SW}). Therefore, in
the case of the $t$-flow, $t$ cannot be explicitly
expressed in terms of $y$
as in the case of
the $x$-flow. This implies that solutions $U(x_0,t)$, $x_0=$const
must have different properties in comparison with (\ref{trace}).

\paragraph{Remark 5.3.} We notice that the subvariety ${W}_2$ of the
generalized Jacobian becomes {\it linear} in rare cases when the curve
$\Gamma$ enjoys some nontrivial
involutions, i.e., when it can be regarded as a covering of an elliptic curve.
(Various examples of the involutions can be found in Belokolos et al [1994].)
In such cases $U(x,t)$ becomes a meromorphic function of its arguments.

\paragraph{The solutions for the Dym equation.}
Now we proceed to the problem of inversion of the reduction (\ref{A-J-odd})
of the Dym equation which is related to the odd order hyperelliptic curve
$\Gamma=\{w^2=R_5(\mu)\}$. As in the
case of the reduction of the SW equation, in
order to describe the function $U(x,t)=\mu_1+\mu_2$,
we first consider an ``excessive'' divisor of 3 points $P_i=(\mu_i,w_i)$,
$i=1,2,3$ on $\Gamma\setminus\{\infty\}$ and the extended equations
\begin{equation}
\label{Generalized-2}
\begin{array}{l}
{\displaystyle
\sum_{i=1}^3\frac{d\mu_i}{2\sqrt{R_5(\mu_i)}}=dy, \quad
\sum_{i=1}^3\frac{\mu_{i}d\mu_{i}}{2\sqrt{R_5(\mu_1)}}=dt, \quad
\sum_{i=1}^3\frac{\mu_i^2 d\mu_{i}}{2\sqrt{R_5(\mu_i)}}=dx, } \\
\\
{\displaystyle
R_5(\mu)=-\kappa\mu(\mu-a_1)\cdots(\mu-a_4), }
\end{array}
\end{equation}
including 2 holomorphic differentials and one differential of
the second kind having
a double pole at $\infty\in\Gamma$. They are linear combinations of the
normalized differentials $\bar{\omega}_1,\bar{\omega}_2,\Omega_{\infty}^{(1)}$
defined in (\ref{norm}) and (\ref{Omega2}).

In contrast to (\ref{A-J2}), equations (\ref{Generalized-2}) describe a
differential of a well defined mapping of the symmetric product
$(\Gamma\setminus\{\infty\})^{(3)}$ to the generalized Jacobian variety
Jac$(\Gamma,\infty)$, the quotient of ${\Bbb C}^3$ by the lattice generated
by {\it four} period vectors of the differentials
$\bar{\omega}_1, \bar{\omega}_2,\Omega_{\infty}^{(1)}$ on $\Gamma$.
Topologically, this variety is a product of the 2 dimensional variety
Jac$(\Gamma)$
and the complex plane ${\Bbb C}$ (see Clebsch [1866], Gavrilov [1999]).

Let us introduce coordinates $z_1,z_2,Z$ by the mapping
\begin{equation}
\label{AJG2}
\sum_{i=1}^3\int_{E_0}^{P_i} \bar{\omega}_1=z_1, \quad
\sum_{i=1}^3\int_{E_0}^{P_i} \bar{\omega}_2=z_2, \quad
\sum_{i=1}^3\int_{E_0}^{P_i}\Omega_{\infty}^{(1)}=Z
\end{equation}
with the basepoint $E_0=(0,0)$ (we cannot choose the basepoint to be $\infty$
as in the previous section, since it is the pole of $\Omega_{\infty}^{(1)}$).
Next, comparing (\ref{norm}), (\ref{Omega2}) with (\ref{Generalized-2}),
we find the following relations
\begin{equation}
\label{norm2}
\begin{array}{l}
{\displaystyle
z_1=d_{11}y+d_{12}t+{\rm const}, \quad z_2=d_{21}y+d_{22}t+{\rm const} } \\%
{\displaystyle
Z=x+h_1'(d_{11}y+d_{12}t)+h_2'(d_{21}y+d_{22}t)+{\rm const}. }
\end{array}
\end{equation}

Like (\ref{AJG}), the mapping (\ref{AJG2}) is invertible in terms of
meromorphic functions.
The inversion problem is solved by means of the following rational
degeneration of the customary theta-function
\begin{equation}
\label{theta-hat}
\widehat{\theta}(z,Z)=Z\theta[\delta](z)+\partial_V\theta[\delta](z),
\end{equation}
where $\partial_V$ is defined in (\ref{trace2}) (compare with the
generalized theta-functions (\ref{gen-theta})). Like (\ref{gen-theta}),
the function (\ref{theta-hat}) enjoys the quasi-periodic property.

Consider again the dissection $\tilde{\Gamma}$ of $\Gamma$ and cut it
along a path joining a point $O$ on the boundary $\partial\Gamma$ to
$\infty$. In the obtained domain we introduce the single-valued function
$$
\widehat{F}(P)=\left( Z-\int_{E_0}^{P}\Omega_{\infty}^{(1)}\right)
{\theta}[\delta]
\left(z-\int_{E_0}^{\infty}{\bar\omega}-\int_{E_0}^{P}{\bar\omega}\right)+
\partial_V{\theta}[\delta]
\left(z-\int_{E_0}^{\infty}{\bar\omega}-\int_{E_0}^{P}{\bar\omega}\right).
$$
Using a modification of Theorem \ref{Riemann} and calculating the logarithmic
differential $\mu(P)\,d\log{\widehat F}(P)$, we obtain
\begin{equation}
\label{3trace-2}
\begin{array}{l}
{\displaystyle
\mu_1+\mu_2+\mu_3={\rm const}
-(Z+\partial_V\log\theta[\delta+\eta_0](z))^2
-\partial_V^2\log\theta[\delta+\eta_0](z) } \\
\\
{\displaystyle
={\rm const} -Z^2-\frac{2Z\partial_V\theta[\delta+\eta_0](z)
-\partial_V^2\theta[\delta+\eta_0](z)}{\theta[\delta+\eta_0](z)},}
\end{array}
\end{equation}
where $\eta_0=(\eta_0'',\eta_0')^T\in\frac 12 {\Bbb Z}^2/{\Bbb
Z}^2$, such that $2\pi i\, \eta_0''+{\bf B}\eta_0'
=\int_{E_0}^{\infty}(\bar{\omega}_1,\bar{\omega}_2)^T$. Now,
similarly to the case of the SW equation, we fix $P_3\equiv E_0$
($\mu_3=0$, $d\mu_3\equiv 0$) in the mapping (\ref{AJG2}). In this
case its image becomes a 2-dimensional nonlinear noncompact
analytic subvariety ${\widehat W}_2\subset{\rm
Jac}(\Gamma,\infty)$. Comparing the third sum in (\ref{AJG2}) and
expression (\ref{AB-trans-2}), we find
\begin{equation}
\label{hatW}
\widehat{W}_2=\{Z+{\rm const}+\partial_V\log\theta[\delta+\eta_0](z)=0\}.
\end{equation}

Finally, taking into account the trace formula (\ref{3trace-2}),
we conclude that the solution of the Dym equation has the form
\begin{equation}
\label{UU}
\begin{array}{l}
{\displaystyle
U(x,t)=\mu_1+\mu_2={\rm const}-\partial_V^2\log\theta[\delta+\eta_0](z), } \\
\\
{\displaystyle
z_1=d_{11}y+d_{12}t+{\rm const}, \quad z_2=d_{21}y+d_{22}t+{\rm const}, }
\end{array}
\end{equation}
where the extra variable $y$ depends on $x,t$ in a transcendental way
according to the constraint (\ref{hatW}) and the expression for $Z$
in (\ref{norm2}). The solution $U(x,t)$ is not meromorphic with respect to
its arguments.

\paragraph{Remark 5.4.} As in the case of SW equation,
the stationary solutions for the Dym equation given in the
previous section can be obtained from (\ref{UU}) by putting
$t=$const. Then $y$ can be identified with the auxiliary variable
$x'$ defined in (\ref{change}) and the condition in (\ref{hatW})
becomes equivalent to the relation (\ref{trans1}) between $x$ and
$x'$. As a result, (\ref{UU}) gives precisely the stationary
solution (\ref{trace2}).

\section{Real bounded stationary 2-phase solutions.}
\setcounter{equation}{0}
In this section we impose reality conditions on the stationary complex
solutions obtained in Section 4.

Let $\sigma$ be the antiholomorphic involution on a hyperelliptic curve
$\Gamma=\{w^2=P(\mu)\}$ of genus $g$.
The part of $\Gamma$ which is invariant with respect to
$\sigma$ is called the real part $\Gamma(\Bbb R)$. On the plane
$\Bbb R^2=(\Re\mu,\Re w)$ it is either the empty set or a union of ovals.
By the Abel--Jacobi mapping, the involution $\sigma$ lifts to Jac($\Gamma$).
By ${\rm Jac}_{\Bbb R}(\Gamma)$ we denote the
real part of Jac($\Gamma$) that is invariant under $\sigma$.
One can show that the elementary symmetric functions of the variables
$\mu_1,\dots,\mu_g$ take real values on ${\rm Jac}_{\Bbb R}(\Gamma)$ and
only there.

\begin{thm} (Comessatti [1924]). Let $s$ be the number of connected
components of
$\Gamma(\Bbb R)$ and $L$ be the number of connected components of
${\rm Jac}_{\Bbb R}(\Gamma)$.
If $s\ne 0$, then $L=2^{s-1}$. If $s=0$, then $L=1$ provided the degree
of $R(\mu)$ is even and $L=2$ in case the degree is odd.
\end{thm}

\paragraph{Shallow water equation.}
Suppose all the roots of the polynomial $R_6(\mu)$ in (\ref{A-J})
arising in the reduction
of the SW equation are real, i.e., $\Gamma(\Bbb R)$ consists of 3
ovals about the segments $[0,e_1]$, $[e_2,e_3]$, and $[e_4,e_5]$.
By Theorem 5.1, ${\rm Jac}_{\Bbb R}(\Gamma)$ has 4 connected components.
They are characterized by the following behavior of $\mu$-variables,
which reflects in different properties of real stationary solutions
$U(x,t_0)$:

\paragraph{Case 1.} The variables $\mu_1,\mu_2$ are real and
$\mu_1\in[a_2,a_3]$, $\mu_2\in[a_4,a_5]$. The sum
$U=\mu_1+\mu_2$ is thus a real quasi-periodic function of $x'$ having no poles
and zeros. The product $\mu_1\mu_2$ has the same properties. Geometrically
this means that the corresponding component of ${\rm Jac}_{\Bbb R}(\Gamma)$
does not intersect the translates $\Theta_-,\Theta_+$, and $\Theta_0$.
In view
of (\ref{change}), $x(x')$ and $x'(x)$ are strictly monotonic real functions.
Therefore the composition $U(x,t_0)=U(x'(x))$ is a quasi-periodic regular
function.

\paragraph{Cases 2,3.} $\mu_1\in[0,a_1]$, whereas $\mu_2\in[a_2,a_3]$ or
$[a_4,a_5]$. The function $U(x')$ has the same properties as
above, whereas $\mu_1\mu_2$ does not blow up, but has zeros. As
found in Section 4, the derivative $dx/dx'$ vanishes with a second
order with respect to $x'$ as one of the $\mu$-variables vanishes.
This happens when the real $x'$-flow on the considered components
of ${\rm Jac}_{\Bbb R}(\Gamma)$ intersects $\Theta_0(\Bbb
R)=\Theta_0 \cap{\rm Jac}_{\Bbb R}(\Gamma)$. If follows that at
this moment the derivative $dU/dx=dU/dx'\cdot dx'/dx$ blows up and
the graph of the function $U(x,t_0)$ has an inflection point with
vertical tendency line.

In addition, when $(\mu_1,\mu_2)=(0,a_2)$ or $(0,a_4)$, i.e., when the
real $x'$-flow passes a half-period on ${\rm Jac}_{\Bbb R}(\Gamma)$,
the function $U(x')$ has an extremum, which implies that the graph of
$U(x,t_0)$ has a cusp. Due to quasi-periodicity of the flow, $U(x,t_0)$ has
an infinite quasi-periodic sequence of cusps.

\paragraph{Case 4.} Now the variables $\mu_1,\mu_2$ are complex conjugated.
Using equations (\ref{A-J}), we show that they cannot reach real axis.
It follows that the product $\mu_1\mu_2$ is always nonzero and
$U(x,t_0)$ is again a quasi-periodic regular function.
\medskip

In a similar  way one can show that when some of the roots of
$R_6(\mu)$ are complex conjugate, the qualitative behavior of the
real solution $U(x,t_0)$ coincides with one of the above four
cases.

In a forthcoming paper  we will consider different singular limits of the
quasi-periodic solutions
when the spectral curve becomes singular and its arithmetic genus drops to
zero.
The solutions are then expressed in terms of purely exponential tau-functions
and, in the real bounded case, they describe an interaction of the two
smooth solitons
or cuspons, or a quasi-periodic train of peakons tending to a periodic one
at infinity.

\section{Peakon-soliton solutions and elliptic billiards.}
\setcounter{equation}{0}
In this section we continue studying degenerate solutions of the Dym equation.
Consider another possible confluence of roots of the polynomial
$R_5(\mu)$ in (\ref{A-J-odd}) by putting
\begin{equation}
\label{limits} a_1=0, \;\; a_2=a_3=b, \;\; a_4=a, \;\; \kappa=1.
\end{equation}
As before, we first analyze stationary solutions ($x$-flow) by
setting $dt=0$. Passing to the new variable $x'$ according to the
change
\begin{equation} \label{change3}
dx=\mu_1\mu_2\, dx',
\end{equation}
from (\ref{A-J-odd}) we get
\begin{equation}
\label{peaksol1}
\begin{array}{l}
{\displaystyle
\frac{d \mu_1}{\mu_1 \sqrt{a-\mu_1}}+\frac{d\mu_2}{\mu_2\sqrt{a-\mu_2}}
= -b dx',} \\
{\displaystyle
\frac{d \mu_1}{(\mu_1-b)\sqrt{a-\mu_1}} +
\frac{d \mu_2}{(\mu_2-b)\sqrt{a-\mu_2}}= 0. }
\end{array} \end{equation}
Let us introduce the following normalized differentials of third kind on
the Riemann surface ${\Bbb P}=\{\xi^2=a-\mu\}$ that have pairs of simple poles
$Q_1^-, Q_1^+$ and $Q_2^-, Q_2^+$ with $\mu=0$ and $\mu=b$ respectively
$$
\Omega_1=\frac{\beta_1\,d\mu}{\mu\xi}, \quad
\Omega_2=\frac{\beta_2\,d\mu}{(\mu-b)\xi}, \qquad
\beta_1=\sqrt{a}, \;\; \beta_2=\sqrt{a-b}.
$$
Then equations (\ref{peaksol1}) give rise to the generalized Abel--Jacobi
equations
\begin{equation}
\label{AJG-2}
\int_{P_0}^{P_1}\Omega_1+\int_{P_0}^{P_2}\Omega_1=z_1, \quad
\int_{P_0}^{P_1}\Omega_2+\int_{P_0}^{P_2}\Omega_2=z_2, \qquad
P_i=(\mu_i,\xi_i),
\end{equation}
\begin{equation}
\label{z12}
z_1=-b\beta_1 x'+z_{10}, \quad z_{10},z_2={\rm const},
\end{equation}
where we put $P_0=(a,0)$.
These describe a well defined mapping of the symmetric product
$({\Bbb P}\setminus\{Q_1^-,Q_1^+,Q_2^-,Q_2^+\})^{(2)}$
to the generalized Jacobian  Jac(${\Bbb P},Q_1^\pm,Q_2^\pm $).

As a result of inversion of (\ref{AJG-2}), one finds
the following expressions for symmetric polynomials of $\mu_1$ and $\mu_2$
\begin{equation}
\label{ps}
\begin{array}{l}
{\displaystyle
\mu_1+\mu_2=U(z_1,z_2)
=\partial^2_{W} \log \tau(z_1,z_2)+\beta_1^2-\beta_2^2 }\\
{\displaystyle
=4(\beta_1^2-\beta_2^2)\frac {\beta_1^2(e^{z_2}+e^{-z_2})
+\beta_2^2(e^{z_1}+e^{-z_1})+2(\beta_1^2-\beta_2^2)}
{\tau^2(z_1,z_2)}+\beta_1^2-\beta_2^2 ,} \\
\\
{\displaystyle
\mu_1 \mu_2=-\frac1{b\beta_1}\partial_{z_1} \partial_{W} \log \tau(z_1,z_2)
\quad {\rm or} \quad
\mu_1\mu_2=4ab \frac{(e^{z_2/2}+e^{-z_2/2})^2}{\tau^2(z_1,z_2)}, }
\end{array}  \end{equation}
where
$$
\partial_{W}=2\beta_1 \frac{\partial}{\partial z_1} +
2\beta_2\frac{\partial}{\partial z_2}, \qquad z_1=-b\beta_1
x'+z_{10}, \quad z_{10},z_2={\rm const}, $$ and $\tau(z_1,z_2)$ is
the 2-dimensional tau-function with $\alpha_1,\alpha_2$ replaced
by $\beta_1,\beta_2$. The latter admits decomposition in the
following sum of one-dimensional tau-functions
\begin{equation}
\label{decomposition}
\begin{array}{l}
{\displaystyle
\tau(z_1,z_2)=e^{z_1/2}\tau(z_2+q/2)-e^{-z_1/2}\tau(z_2-q/2), } \\%
{\displaystyle
\tau(z_2)=e^{z_2/2}+e^{-z_2/2}, \quad
q=\log\left( \frac{\beta_1-\beta_2}{\beta_1+\beta_2}\right)^2. }
\end{array}
\end{equation}
Lastly, by using the second expression in (\ref{ps}) and the relation between
$z_1$ and $x'$ in (\ref{z12}), we find
$$
x=\int \mu_1\mu_2\, dx'=\partial_{W} \log \tau(z_1,z_2)+{\rm const}
$$
or, in view of decomposition (\ref{decomposition}),
\begin{equation}
\label{x-x}
\begin{array}{l}
{\displaystyle
x(x')=\beta_1 \frac{e^{z_1/2}\tau(z_2+q/2)+e^{-z_1/2}\tau(z_2-q/2)}
{e^{z_1/2}\tau(z_2+q/2)-e^{-z_1/2}\tau(z_2-q/2) } } \\
{\displaystyle
+\beta_2 \frac{e^{z_1/2}\partial_{z_2}\tau(z_2+q/2)
-e^{-z_1/2}\partial_{z_2}\tau(z_2-q/2)}
{e^{z_1/2}\tau(z_2+q/2)-e^{-z_1/2}\tau(z_2-q/2)}+{\rm const}, } \\%
\\
z_1=-b\beta_1 x'+z_{10}, \quad z_2={\rm const}.
\end{array}
\end{equation}

\paragraph{Remark 8.1.} As mentioned in Remark 4.1, equations (\ref{A-J-odd})
with $dt=0$ describing the quasiperiodic stationary solutions have
the same structure as quadratures for the geodesic motion on an
triaxial ellipsoid ${\cal E}$ (more generally, a quadric) in $\Bbb
R^3$. Parameter $x$ plays the role of length of a geodesic. Under
the limit (\ref{limits}) one of the semiaxes of ${\cal E}$ tends
to zero whereas the geodesic motion passes to the {\it asymptotic
billiard motion} inside the ellipse $$ \bar{\cal
E}=\left\{X_1^2/a+X_2^2/b=1\right\}\subset {\Bbb R}^2=(X_1,X_2).
$$ Geodesics on ${\cal E}$ transform to straight line segments
passing through a focus of the ellipse between each subsequent
elastic reflections (impacts) along $\bar {\cal E}$. As
$x\to\pm\infty$, the billiard motion tends to oscillations along
the bigger axis of the ellipse. Now the variables $\mu_1,\mu_2$
play the role of elliptic coordinates in ${\Bbb R}^2$ such that $$
X_1^2=\frac{(a-\mu_1)(a-\mu_2)}{a-b}, \quad
X_2^2=\frac{(b-\mu_1)(b-\mu_2)}{b-a}. $$ Along $\bar {\cal E}$ one
of the variables equals zero.

It follows that equations (\ref{peaksol1}) can be regarded as
describing the straight line motion of a point mass inside $\bar
{\cal E}$. When the point meets the ellipse, one of the
$\mu$-variables, say $\mu_1$, vanishes, and the corresponding
point $P_1=(\mu_1,\sqrt{R(\mu_1)})$ on the Riemann surface $\Bbb
P$ coincides with one of the poles $Q_1^-,Q_1^+$ of the
differential $\Omega_1$. Then, as follows from the mapping
(\ref{AJG-2}), $z_1$ and $x'$ become infinite.

On the other hand,
as $x',z_1\to\pm\infty$, the second expression in (\ref{ps}) vanishes,
whereas the first one has finite limits giving the values of
$\mu_2$ at the subsequent impact points. The variable $z_2$ plays the role
of a constant phase defining position of the segment between the points.

According to (\ref{x-x}), (\ref{ps})
\begin{equation}
\label{ends-x}
\begin{array}{l}
{\displaystyle
x(-\infty,z_2)=-\beta_1+\beta_2\frac{\partial_{z_2}\tau(z_2-q/2)}{\tau(z_2-q/2)}
+c_x,} \\%
{\displaystyle
x(\infty,z_2)=\beta_1+\beta_2\frac{\partial_{z_2}\tau(z_2+q/2)}{\tau(z_2+q/2)}
+c_x, \quad c_x={\rm const}. }
\end{array}
\end{equation}
and
\begin{equation}
\label{ends-mu}
\begin{array}{l}
{\displaystyle
U(-\infty,z_2)= \frac{4\beta_2^2 (\beta_1^2-\beta_2^2)}
{ (\beta_1-\beta_2)^2 e^{-z_2}+(\beta_1+\beta_2)^2 e^{z_2}
+2(\beta_1^2-\beta_2^2) } +\beta_1^2-\beta_2^2,} \\%
{\displaystyle
U(\infty,z_2)= \frac{4\beta_2^2 (\beta_1^2-\beta_2^2)}
{ (\beta_1-\beta_2)^2 e^{z_2}+(\beta_1+\beta_2)^2 e^{-z_2}
+2(\beta_1^2-\beta_2^2) } +\beta_1^2-\beta_2^2. }
\end{array}
\end{equation}

Notice that $x(\infty,z_2)$ and $U(\infty,z_2) $ have the same value as
$x(-\infty,z_2+q)+2\beta_1$ and $U(-\infty,z_2+q)$ respectively.
All this results in the following algebro-geometrical description:
{\it As the point mass inside $\bar{\cal E}$ moves from one impact to the
next one,
the point $P_1$ on $\Bbb P$ moves from the pole $Q_1^-$ to $Q_1^+$.
At the moment of impact, $P_1$ jumps from $Q_1^+$ back to $Q_1^-$,
whereas the phase $z_2$ in (\ref{ps}) increases by $q$. Then the story repeats.}

Using this property, by induction, from (\ref{ends-mu}) the elliptic
coordinates of the whole sequence of impact points are found in form
\begin{equation}
\label{impact}
\begin{array}{l}
{\displaystyle
\mu_1=0, \quad \mu_2=\frac{4\beta_2^2 (\beta_1^2-\beta_2^2)}
{ (\beta_1-\beta_2)^2 e^{z_{2N}}+(\beta_1+\beta_2)^2 e^{-z_{2N}}
+2(\beta_1^2-\beta_2^2) } +\beta_1^2-\beta_2^2,} \\
{\displaystyle
z_{2N}=z_{20}+Nq, } \end{array}
\end{equation}
$N\in{\Bbb N}$ being the number of impact and the constant
$z_{20}$ is the same for all the segments of the billiard
trajectory.

In addition, from (\ref{ends-x}) we find the length of the $N$-th
segment of the billiard trajectory in form
\begin{equation}
\label{lenght}
x(\infty,z_{2N})-x(-\infty,z_{2N})=2\beta_1+
2\beta_2 \frac{e^{q/2}-e^{-q/2}}{\exp(z_{2N})-\exp(-z_{2N})+e^{q/2}-e^{-q/2}},
\end{equation}
$z_{20}$ being the same as in (\ref{impact}).

According to the trace formula $U(x,t_0)=\mu_1+\mu_2$, expressions
(\ref{ps}), (\ref{x-x}) provide us stationary {\it peakon}
solutions of Dym equation in a parametric form. Here the phase
$z_2$ must be regarded as a certain function of $t_0$. Namely, the
expressions describe one piece of the profile $U(x,t_0)$
corresponding to trajectory of the point mass between subsequent
impacts. The other pieces are obtained by changing the phase $z_2$
in (\ref{ps}), (\ref{x-x}) by $q$ and adding $2\beta_1$ to $x$.
The pieces are glued at {\it peak points}, where the spatial
derivative of $U$ changes sign and which correspond to impacts in
the billiard problem. The profile $U(x,t_0)$ thus consists of an
infinite sequence of peaks and knots between them. For this reason
we call this solution {\it the soliton-peakon solution}. Notice
that, in contrast to exponentials profiles of the functions
$U(x')$, $x(x')$, any piece of $U(x,t_0)$ has {\it quadratic}
profile, as will be explained below.

The heights $U_N$ of $N$-th peak is given by (\ref{impact}). The
distance between subsequent peaks along $x$-axis is a
quasiperiodic function of $N$ given by (\ref{lenght}).

We emphasize that, in contrast to the peakon traveling wave
solution considered in Section 3, now the $x$-distance between
subsequent peaks is not constant. However, as seen from
(\ref{lenght}), for $N\to\pm\infty$ it tends to $2\beta_1$, the
doubled bigger semi-axis of the ellipse, whereas the pieces tend
to identical ones corresponding to periodic billiard motion along
$X_1$-axis.
\medskip

\paragraph{Remark 8.2.} Expressions (\ref{ps}), (\ref{x-x}) describe an
asymptotic
motion of an elliptic as well as a hyperbolic billiard. In
the first case the initial phase $z_{10}$ is pure imaginary whereas
in the second case it is real. According to the trace formula, the hyperbolic
billiard corresponds to unbounded stationary solutions of HD equation,
which is out of interest of this paper.

\paragraph{Remark 8.3}. The above considerations can be extended to
multi-dimensional case. Namely, following similar approach one can
consider genus $g$ solution of HD equation described by equations
(\ref{A-J-g}), then its asymptotic stationary limit which is
related to a generalized Abel--Jacobi mapping including $g$
meromorphic differentials of 3rd kind. Consider a billiard inside
a $g$-dimensional ellipsoid. Then such a limit solution
corresponds to asymptotic billiard trajectories that intersect
$g-1$ focal quadrics of the ellipsoid between any subsequent
impacts. The resulting stationary solution $U(x,t_0)$ consists of
an infinite series of peaks and between each subsequent peaks
there are $g-1$ knots.
\medskip

In order to study time-dependent soliton-peakon solutions, we
consider the system (\ref{A-J-odd}) under the limits
(\ref{limits}) without changing the variable $x$. As a result,  we
arrive at
\begin{equation}
\label{abel3}
\begin{array}{l}
{\displaystyle
\frac{d\mu_{1}}{2(\mu_1-b)\sqrt{a-\mu_1}}+
\frac{d\mu_{2}}{2(\mu_2-b)\sqrt{a-\mu_2}}=dt, } \\%
{\displaystyle
\frac{d\mu_{1}}{2\sqrt{a-\mu_1}}+
\frac{d\mu_{2}}{2\sqrt{a-\mu_2}}=dx-b\,dt. }
\end{array}
\end{equation}

The latter equations include one differential of third kind
$\Omega_2=\frac{\beta_2\,d\mu}{(\mu-b)\xi}$ having simple poles
$Q_2^-$, $Q_2^+$ and one differential of
second kind $\Omega_{\infty}^{(1)}$ having a double pole at infinity
on the Riemann surface ${\Bbb P}=\{\xi^2=a-\mu\}$. Consider the mapping
\begin{equation}
\label{map2}
\int_{P_0}^{P_1}\Omega_{\infty}^{(1)}
+\int_{P_0}^{P_2}\Omega_{\infty}^{(1)}=z_1, \quad
\int_{P_0}^{P_1}\Omega_2+\int_{P_0}^{P_2}\Omega_2=z_2, \qquad
P_i=(\mu_i,\xi_i),
\end{equation}
where, $z_1=\beta_2 t+Z_{10}$, $z_2=x+bt+z_{20}$, $z_{10},z_{20}=$const
and, as above, $P_0=(a,0)$. Integrating it explicitly, we obtain
\begin{equation}
\label{finite}
\frac{(\xi_1-\beta_2)(\xi_2-\beta_2)}
{(\xi_1+\beta_2)(\xi_2+\beta_2)}=e^{z_2}, \quad \xi_1+\xi_2=z_1.
\end{equation}
Inverting these relations yields the following formal solution for the
HD equation
\begin{equation}
\label{direct_solution}
\begin{array}{l}
{\displaystyle
U(x,t)=\mu_1+\mu_2=(\xi_1+\xi_2)^2-2\xi_1\xi_2
=z_1^2-2\beta_2(\beta_2-z_1)\frac{e^{z_2/2}+e^{-z_2/2}}{e^{z_2/2}-e^{-z_2/2}},}
\\
{\displaystyle
z_1=x+bt+z_{10},  \quad z_2=\beta_2 t+z_{20}, \qquad
z_{10}, z_{20}={\rm const} .}
\end{array}
\end{equation}
It is seen that $U$ depends on $x$ rationally (quadratically, as already
mentioned above) and $U$ is unlimited as $x\to\pm\infty$.

However, this solution does not take into account the reflection
phenomenon described above: when the variable $\mu_1$ vanishes, the
corresponding point $P_1\in{\Bbb P}$ jumps from the pole $Q_2^+$ of
the differential $\Omega_2$ to $Q_2^-$. According to mapping (\ref{map2}),
this results in changing the phases $z_1,z_2$ in (\ref{direct_solution}) by the
constants
$$
\int_{Q_2^-}^{Q_2^+}\Omega_{\infty}^{(1)}=2\beta_1,\quad {\rm respectively}
\quad
 q=\int_{Q_2^-}^{Q_2^+}\Omega_2=2\log\frac{\beta_1-\beta_2}{\beta_1+\beta_2},
$$
the latter being already defined in (\ref{decomposition}).
It follows that the actual solution $U(x,t)$ to HD equation consists
of an infinite number of pieces described by (\ref{direct_solution}) with
\begin{equation}
\label{series}
z_1=x+bt+2\beta_1 N+z_{10}, \quad z_2=\beta_2 t+Nq+z_{20}, \quad N\in\Bbb Z
\end{equation}
and glued along {\it peak lines} $\{x=q_N(t)\}$ in the plane $(x,t)$, where
for a fixed time $t$, the function $q_N(t)$ gives $x$-coordinate of $N$-th
peakon.
Now if we assume $z_{20}$ to be imaginary and $z_{10}$ real, the series of
pieces will provide a real bounded peakon solution.

The functions $q_N(t)$ can be found as follows. Along the peak
lines one of the variables $\mu$, say $\mu_1$, vanishes implying
$d\mu_1=0$, $\xi_1\equiv\beta_2$. Putting this into
(\ref{finite}), we find $$
\frac{\xi_2-\beta_2}{\xi_2+\beta_2}=e^{z_2-q/2}, \quad
\beta_2+\xi_2=z_1. $$ Substituting here (\ref{series}), putting
$x=q_N(t)$, and eliminating $\xi_2$, we obtain the sought
expression $$
q_N(t)=-bt-z_{10}-2\beta_1N+\beta_1+\beta_2\frac{1+e^{z_2-q/2}}{1-e^{z_2-q/2
}}, \quad z_2=\beta_2 t+Nq+z_{20}. $$ It follows that for $|t|>>1$
any peak moves with constant velocity $-b$, and as $t$ evolves
from $-\infty$ to $\infty$ the peaks undergo the phase shift $x\to
x-2\beta_2$.

\newpage

\section*{Bibliography.}

\begin{description}

\item Abenda, S. and Fedorov.Yu [1999], On the weak Kowalewski--Painlev\'e
property for hyperelliptically separable systems,
{\it Acta Appl. Math.\/} (to appear).

\item  Ablowitz, M.J. and Segur, H [1981],
Solitons and the Inverse Scattering Transform, SIAM, Philadelphia.

\item Alber, M.S. and  Alber, S.J. [1985], Hamiltonian formalism for
finite-zone solutions of integrable equations,
{\it C.R. Acad. Sc. Paris\/} {\bf 301}, 777-781.

\item Alber, M.S., R. Camassa, D.D. Holm and J.E. Marsden [1994],
The geometry of peaked solitons and billiard solutions of a class of
integrable pde's, {\it Lett. Math. Phys.\/} {\bf 32} 137-151.

\item Alber, M.S., Camassa, R., Holm, D.D., and Marsden, J.E. [1995], On
the link between umbilic geodesics and soliton solutions of nonlinear PDE's,
{\it Proc. Roy. Soc} {\bf 450} 677-692.

\item Alber, M.S., R. Camassa, Y. Fedorov, D.D. Holm and J.E. Marsden [1999],
On Billiard Solutions of Nonlinear PDE's, {\it Phys. Lett. A} {\bf
264} 171--178.

\item Alber, M.S.,  and C. Miller [1999], On Peakon Solutions of the
Shallow Water Equation, {\it Appl.Math. Lett.} (to appear).

\item Alber, M.S., Camassa, R., Fedorov, Yu., Holm, D.D., and Marsden J.E.
[1999], The geometry of new classes of weak billiard solutions of
nonlinear PDE's. (subm.)

\item Belokolos, E.D., A.I. Bobenko, V.Z. Enol'sii, A.R. Its, and
V.B. Matveev [1994] {\it Algebro-Geometric Approach to Nonlinear
Integrable Equations.\/} Springer-Verlag series in Nonlinear Dynamics.

\item Beals, R., D.H. Sattinger, J. Szmigielski [1999],
Multi-peakons and a theorem of Stietjes, {Inverse Problems} {\bf
15} L1--L4.

\item Beals, R., D.H. Sattinger, J. Szmigielski [2000],
Multipeakons and the Classical Moment, {Advances in Mathematics}
(to appear)

\item Camassa, R. and Holm, D.D. [1993], An integrable
shallow water equation with peaked solitons,
{\it Phys. Rev. Lett.\/}, {\bf 71}, 1661-1664.

\item Camassa, R., Holm, D.D., and Hyman, J.M. [1994],
A new integrable shallow water equation.
{\it Adv. Appl. Mech.}, {\bf 31}, 1--33.

\item Cewen, C. [1990], Stationary Harry-Dym's equation and its
relation with geodesics on
ellipsoid, {\it Acta Math. Sinica} {\bf 6}, 35--41.

\item  Clebsch, A., Gordan, P. [1866], Theorie der abelschen Funktionen.
Teubner, Leipzig

\item Comessatti, A. and  Sulle variet\'a abeliane reale I.
{\it Ann. Math. pure Appl.}, {\bf 2}, (1924) 67--106.

\item Dmitrieva, L.A. [1993a] Finite-gap solutions of the Harry Dym equation,
{\it Phys.Lett.A} {\bf 182} (2) 65--70.

\item Dmitrieva, L.A. [1993b], The higher -times approach to multisoliton solutions of the
Harry Dym equation, {\it J.Phys.A} {\bf 26} 6005--6020.

\item Dubrovin, B. [1981], Theta functions and nonlinear equations. {\it
Russ. Math. Surveys\/}, {\bf 36}, 11--92.

\item Dubrovin, B.A., Novikov S.P., Krichiver, I.M. [1985]
{\it Integrable Systems. I.
Itogi Nauki i Tekhniki. Sovr.Probl.Mat. Fund.Naprav. 4,
VINITI,} Moscow. English transl.: Encyclopaedia of Math.Sciences,
Vol. 4, Springer-Verlag, Berlin 1989.

\item  Ercolani, N. [1989], Generalized theta functions and
homoclinic varieties, {\it Proc.
Symp. Pure Appl. Math.\/}, {\bf 49}, 87--100.

\item Fedorov, Yu. [1999], Classical integrable systems and billiards
related to generalized Jacobians, {\it Acta Appl. Math.}, {\bf 55}, 3,
151--201

\item Gagnon, L., Harnad, J., Hurtubise, J. and Winternitz, P. [1985],
Abelian integrals and the reduction method for an integrable
Hamiltonian system, {\it J.Math.Phys.} {\bf 26}, 1605--1612.

\item Gavrilov, L. [1999], Generalized Jacobians of spectral curves
and completely integrable systems,  {\it Math.Z.} (to appear).

\item Gunning, R. [1972], Lectures on Riemann Surfaces. Jacobi varieties.
Princeton University Press.

\item Hunter, J.K., and  Zheng, Y.X. [1994], On a completely
integrable nonlinear hyperbolic variational equation,
{\it Physica D} {\bf 79}, 361--386.

\item Li, Y.A. and Olver P.J. [1998], Convergence of solitary-wave solutions
in a perturbed bi-Hamiltonian dynamical system. {\it Discrete and
continuous dynamical systems}, {\bf 4}, 159--191.

\item  Markushevich, A. I. [1977], Theory of Functions of a Complex Variable,
Chelsea Publishing Company: New York.

\item McKean, H.P. and A. Constantin [1999], A Shallow Water Equation on
the Circle, {\it Comm.Pure Appl.Math.} {\bf Vol LII} 949--982.

\item Mumford, D. [1983], Tata Lectures on Theta II, Birkhauser-Verlag.

\item Novikov, D.P. [1999], Algebraic geometric solutions of the Harry Dym Equations, {\it
Siberian Math. J.} {\bf 40} 136--140.

\item Previato E. [1985],
Hyperelliptic quasi-periodic and soliton solutions of the
nonlinear Schr\"odinger equation {\it Duke Math. J.} {\bf 52}, I,
329--377.

\item Vanhaecke, P. [1995], Integrable systems and symmetric products
of algebraic curves, {\it Math. Z.} {\bf 40} 143--172.

\item Weierstrass, K. [1878],
\"Uber die geod\"atischen Linien auf dem dreiachsigen
    Ellipsoid, Mathematische Werke I, 257--266.

\item Wadati M., Konno, H., Ichikawa, Y.H. [1979], {\it J. Phys. Soc. Japan} {\bf 47}, 1698.
\end{description}
\end{document}